\documentclass[traditabstract,longauth]{aa}

\usepackage{gensymb}
\usepackage{amsmath}
\usepackage{graphicx}
\usepackage{txfonts}
\usepackage{natbib}
\usepackage{color}
\usepackage{multirow}
\usepackage{subfigure}
\usepackage{lscape}
\usepackage{latexsym}
\bibpunct{(}{)}{;}{a}{}{,} 
\usepackage[usenames,dvipsnames,svgnames,table]{xcolor}
\usepackage[breaklinks, colorlinks, citecolor=CornflowerBlue]{hyperref}
\hypersetup{draft}
\usepackage{url}
\usepackage{longtable}
\usepackage{bm}

\newcommand{\ha}{H$\alpha$}

\newcommand{\Msun} {$\mathrm{M}_{\sun}$}

\begin{document} 

   \title{A steep mass transition for bar-driven ISM structuring revealed by PHANGS-JWST}
   \titlerunning{Mass-dependent bar-driven ISM evolution}
\author{
   Eric~Emsellem\inst{\ref{eso},\ref{lyon}} \and 
   Jessica Sutter\inst{\ref{whitman}} \and
   Ryan Chown\inst{\ref{Alg}} \and
   Adam Leroy\inst{\ref{Ohio},\ref{CfC}} \and
   Debosmita Pathak\inst{\ref{Ohio},\ref{CfC},\ref{IPAC}} \and
   Eva Schinnerer\inst{\ref{MPIA}} \and
   Thomas G. Williams\inst{\ref{JBCA}} \and
   Karin Sandstrom\inst{\ref{UC}} \and
   Oscar Agertz\inst{\ref{lund}} \and
   Francesco Belfiore\inst{\ref{eso}} \and
   Amelia Fraser-McKelvie\inst{\ref{eso}} \and
   Kirsten L. Larson\inst{\ref{STSciESA}} \and
   Janice Lee\inst{\ref{STSci},\ref{Steward},\ref{Gemini}} \and
   Jay~Gonz\'alez~Lobos\inst{\ref{MPIA}} \and
   Sharon Meidt\inst{\ref{Gent}} \and
   Miguel~Querejeta\inst{\ref{OAN}} \and
   Sophia~K.~Stuber\inst{\ref{NAOJ},\ref{MPIfR}} \and 
   Pierrick Verwilghen\inst{\ref{eso}} \and 
   Oleg~V.~Egorov\inst{\ref{ARZ}} \and
   Damian R. Gleis\inst{\ref{MPIA},\ref{UHD}} \and
   Jonathan~D.~Henshaw\inst{\ref{MPIA}} \and
   Justus Neumann\inst{\ref{MPIA}} \and
   Yixian Cao\inst{\ref{MPE}} \and
   Daniel~A.~Dale\inst{\ref{uwyo}} \and
   Daizhong Liu\inst{\ref{PMO}} \and
   Elias K. Oakes\inst{\ref{uconn}} \and
   Marina~Ruiz-Garc\'ia\inst{\ref{OAN},\ref{UCM}}\and
   Dave Thilker\inst{\ref{JHU}} \and
   Elizabeth Watkins\inst{\ref{JB}}
   }
   \institute{European Southern Observatory, Karl-Schwarzschild-Stra{\ss}e 2, 85748 Garching, Germany\label{eso} \\
        \email{eric.emsellem@eso.org}
        \and Univ Lyon, Univ Lyon1, ENS de Lyon, CNRS, Centre de Recherche Astrophysique de Lyon UMR5574, F-69230 Saint-Genis-Laval France\label{lyon}
        \and Whitman College, 345 Boyer Avenue, Walla Walla, WA 99362, USA\label{whitman}
        \and Faculty of Computer Science \& Technology, Algoma University, Sault Ste. Marie, ON P6A 2G4, Canada\label{Alg}
        \and Department of Astronomy, Ohio State University, 180 W. 18th Ave, Columbus, OH 43210, USA\label{Ohio}
        \and Center for Cosmology and Astroparticle Physics, 191 West Woodruff Avenue, Columbus, OH 43210, USA\label{CfC}
        \and IPAC, California Institute of Technology, 1200 E. California Blvd, Pasadena, CA 91125, USA\label{IPAC}
        \and Max-Planck-Institut f\"{u}r Astronomie, K\"{o}nigstuhl 17, D-69117, Heidelberg, Germany\label{MPIA}
        \and UK ALMA Regional Centre Node, Jodrell Bank Centre for Astrophysics, Department of Physics and Astronomy, The University of Manchester, Oxford Road, Manchester M13 9PL, UK\label{JBCA}
        \and Department of Astronomy \& Astrophysics, University of California, San Diego, 9500 Gilman Dr., La Jolla, CA 92093, USA\label{UC}
        \and Division of Astrophysics, Department of Physics, Lund University, Box 118, SE-221 00 Lund, Sweden\label{lund}
        \and AURA for the European Space Agency (ESA), Space Telescope Science Institute, 3700 San Martin Drive, Baltimore, MD 21218, USA\label{STSciESA}
        \and Space Telescope Science Institute, 3700 San Martin Drive, Baltimore, MD 21218, USA\label{STSci}
        \and Steward Observatory, University of Arizona, Tucson, AZ 85721, USA\label{Steward}
        \and 30 Gemini Observatory/NSF’s NOIRLab, 950 N. Cherry Avenue, Tucson, AZ, 85719, USA\label{Gemini}
        \and Department of Physics \& Astronomy, Universiteit Gent, Krijgslaan 281 S9, B-9000 Gent, Belgium\label{Gent}
        \and Observatorio Astronómico Nacional (IGN), C/ Alfonso XII, 3, E-28014 Madrid, Spain\label{OAN}
        \and National Astronomical Observatory of Japan, 2-21-1 Osawa, Mitaka, Tokyo 181-8588, Japan\label{NAOJ}
        \and Max Planck Institute for Radio Astronomy, Auf dem Hügel 69, 53121 Bonn, Germany\label{MPIfR}
        \and Astronomisches Rechen-Institut, Zentrum f\"ur Astronomie der Universit\"at Heidelberg, M\"onchhofstraße 12-14, D-69120 Heidelberg, Germany\label{ARZ}
        \and Universit\"{a}t Heidelberg, Department of Physics and Astronomy, Im Neuenheimer Feld 226, D-69120 Heidelberg, Germany\label{UHD}
        \and Max-Planck-Institut f\"{u}r extraterrestrische Physik, Giessenbachstra{\ss}e 1, D-85748 Garching, Germany\label{MPE}
        \and Department of Physics \& Astronomy, University of Wyoming, Laramie, WY 82071, USA\label{uwyo}
        \and Purple Mountain Observatory, Chinese Academy of Sciences, 10 Yuanhua Road, Nanjing 210023, China\label{PMO}
        \and Department of Physics, University of Connecticut, 196A Auditorium Road, Storrs, CT 06269, USA\label{uconn}
        \and Facultad de Ciencias F\'{\i}sicas, Pl. de Ciencias 1, Universidad Complutense de Madrid, E-28040 Madrid, Spain\label{UCM}
        \and Department of Physics and Astronomy, The Johns Hopkins University, Baltimore, MD 21218, USA\label{JHU}
        \and Jodrell Bank Centre for Astrophysics, Department of Physics and Astronomy, The University of Manchester, Oxford Road, Manchester, M13 9PL, UK \label{JB}
        }
   \authorrunning{Emsellem et al.}

   \date{Received April 27, 2026; accepted June 26, 2026}

  \abstract{Galactic bars are thought to play a critical role in the secular evolution of their hosts by, e.g., reorganising the interstellar medium (ISM). We use a sample of 57 star-forming disc galaxies observed with JWST at 3 and 7.7~$\mu$m to probe how the spatial distribution of Polycyclic Aromatic Hydrocarbons (PAH) emission as a structural marker of the cold ISM might depend on galaxy stellar mass and the presence of a bar. We find evidence for a "watershed" at a stellar mass of $10^{10}$~\Msun, marking a fundamental transition in the bar-driven distribution of PAH emission. This confirms trends previously predicted by numerical simulations and observed via ionised gas or UV light. While lower-mass galaxies exhibit a disordered and clumpy distribution of PAH emission regardless of bar presence, higher-mass barred hosts display well-structured dynamical features traced by PAH emission with significant gas reservoirs (e.g., discs and rings) within the central 15\% of the bar radius (R$_b$). Furthermore, we observe a systematic depletion of PAH emission within the [0.2–0.8]\,R$_b$ range in barred systems with stellar masses above $10^{10}$~\Msun. Such central discs, rings, and associated radial dips ("bar deserts") appear to be a mass-dependent phenomenon: ubiquitous in massive galaxies but mostly absent in their lower-mass counterparts. In contrast to the structured features in massive hosts, the disorganised ISM in lower-mass galaxies masks the commonly observed bar-driven signatures. This suggests that tracer selection and dust obscuration may significantly bias observed bar fractions. Our study underlines the existence of two regimes of secular evolution, with different impacts and observability of bar-driven processes: it reaffirms the role of bars as primary drivers of rapid secular evolution in galaxies above $10^{10}$\,\Msun\, while the impact of bars is significantly reduced or delayed below this threshold. It further underscores the need to critically account for these processes when modelling galaxy evolution in cosmological simulations.
  }
\maketitle

\section{Introduction}
About half to two-thirds of massive disc galaxies in the nearby universe host weak or strong bars \citep{Eskridge+2000, Menendez-Delmestre+2007, Sheth+2008, Erwin2018} that are thought to actively shape the spatial distribution of the interstellar medium (ISM), impacting its ability to form stars \citep[e.g.,][]{Lynden-Bell+1972, Weinberg1985, Knapen1995, Sakamoto+1999, Sheth+2005, Regan+2006, Schultheis+2025}. Observational signatures of their evolutionary impact have been probed through photometric and spectroscopic studies, often using morphological, kinematic and gas abundance tracers \citep[e.g.,][]{Seidel+2015, Vera+2016}. Several works \citep[including][]{Carles+2016, Erwin+2017, Erwin2018, Erwin2019, FraserMcKelvie+2020, Mukundan+2025} have now convincingly demonstrated the existence of statistical trends that depend on host stellar mass.

In particular, multiple lines of evidence suggest that a stellar mass of $\sim 10^{10}$~\Msun\ marks a transition above which barred galaxies show notably different properties, including, e.g.: the absolute (and relative) sizes of bars \citep{Erwin2019, Erwin2024}, the prevalence of boxy-peanut-shaped (bar-driven) bulges and bars with "Peak-Shoulders" radial profiles \citep{Erwin+2017, Erwin2023}, the presence of central stellar or molecular rings \citep{Comeron+2010, Diaz-Garcia+2019, Stuber+2023, Gleis+2026}, the two-dimensional distribution of ionised gas \citep{FraserMcKelvie+2020}, the fraction of galaxies that host bars \citep{Sheth+2008} and the fraction of bars with secondary inner bars \citep{Erwin2024}. 
Below a stellar mass of about $10^{10}$~\Msun, galaxies seem to host bars with sizes that show a very mild dependence on mass \citep{Erwin2019, Erwin2024}, whereas above that threshold, the average bar size increases strongly with mass. With a set of almost 700 galaxies observed as part of the MaNGA two-dimensional spectroscopic survey \citep{Bundy+2015}, \cite{FraserMcKelvie+2020} revealed that galaxies with stellar masses lower than $10^{10}$~\Msun\ exhibit \ha\ emission distributed along their bars, while only galaxies above that mass threshold seem to show the presence of \ha\ emission confined to central rings or at the end of the bars. This result confirms previous trends detected in smaller samples \citep{Herrera+2015} from infrared photometry and expand on studies conducted in the near- and far-UV ranges \citep{Diaz-Garcia+2019} via imaging of hundreds of barred galaxies.

Recent numerical simulations of isolated main-sequence star-forming disc galaxies with stellar masses ($\log({\rm M}_{\star} / {\rm M}_{\odot})$) between 9.5 and 11 have reproduced an observed trend in the distribution of gas and star-forming regions within bars \citep{Verwilghen+24,Verwilghen+25} that seems to echo these findings \citep{Herrera+2015, Diaz-Garcia+2019, FraserMcKelvie+2020}. This strongly suggests that a stellar mass transition around $10^{10}$~\Msun\ derives from the complex interplay between gravity (depth of the potential well) and stellar feedback. There is a mass-dependent, continuous trend in the ability of feedback to perturb gas flows organised by stellar bars, and the stellar mass threshold marks the transition between gravity- and feedback-dominated regimes \citep{Verwilghen+25}. Other main findings from \cite{Verwilghen+25} include the systematic emergence of a depressed gas-density region and strong deviations from log-normal density probability distributions within the bar region for the more massive galaxies. The latter connects with the slow but steady build-up of a massive central gaseous and stellar inner disc \citep[see e.g.,][]{Schultheis+2025} on timescales of only a few bar rotations ($\sim 0.5 - 1$~Gyr). In galaxies less massive than the mass threshold, those timescales increase dramatically, amounting to many gigayears of evolution, which significantly lowers the probability of the emergence of such galactic structures accordingly.

In this paper, we use the advent of high-resolution JWST near- and mid-infrared imaging of a sample of nearby, moderately inclined, star-forming main-sequence galaxies with Polycyclic Aromatic Hydrocarbons (PAH) emission as a tracer of the ISM distribution \citep{Leroy+2023, Chown+2025a}. We thus ask whether we can confirm such a mass-dependent change in structure.
In Section~\ref{sec:data}, we briefly describe our dataset and how it was processed. In Section~\ref{sec:res}, we provide the main stellar-mass-driven trends we detect, followed by a discussion in Section~\ref{sec:disc} and concluding remarks in Section~\ref{sec:conc}.

\section{Data}
\label{sec:data}

\subsection{PHANGS JWST imaging}
\label{sec:imaging}
We use a set of 72 star-forming main-sequence galaxies observed as part of the PHANGS-JWST Cycle 1 \citep[GO 2107, PI: J. Lee;][]{Lee2023, Williams2024} and Cycle 2 \citep[GO 3707, PI: A. Leroy;][]{Chown+2025a} Treasury Programmes. Except for NGC\,1808, these 72 galaxies are all part of the initial PHANGS-ALMA galaxy sample \citep{Leroy2021} surveying a set of nearby ($D \lesssim 20$~Mpc) star-forming main sequence galaxies more massive than $10^{9.5}$~\Msun\ at moderate inclination ($i < 70\degree$). All targets were observed in at least 6 bands. All the selected galaxies have F300M and F335M from NIRCam, and F770W and F2100W from MIRI \citep[see Table~3 in][]{Chown+2025a}. In this paper, we focus exclusively on the 3 and 7.7\,$\mu$m images (NIRCam F300M and MIRI F770W, respectively).

All data were reduced as described in \cite{Chown+2025b} using a dedicated pipeline developed for the PHANGS survey \citep[{\tt pjpipe}\footnote{https://pjpipe.readthedocs.io/en/latest/};][]{Williams2024}.
Throughout this paper, we focus exclusively on two tracers: stellar continuum using F300M, and 7.7~$\mu$m PAH intensity using starlight-subtracted F770W \citep[e.g.,][]{Engelbracht+2005, Engelbracht+2008}. The starlight-subtracted images, F770W$_{\rm ss}$ (sometimes referred to as 'PAH emission' in the following sections; see Sect.~\ref{sec:f77ss} for details), are extracted via the prescription used by \cite{Sutter+2024} and applied elsewhere \citep{Chown+2025a, Sutter+2025, Bazzi+2026}. We scale the 3~$\mu$m image with      ${\rm F770W}_{\rm ss} = {\rm F770W} - 0.22 \times {\rm F300M}$.

All F770W$_{\rm ss}$ and F300M images were deprojected using the tabulated values of their position angles and inclination (see Table~\ref{tab:sample}). We proceeded with a single regridding step using the {\tt reproject\_adaptive} routine of the {\tt astropy} affiliated {\tt reproject} Python package\footnote{https://pypi.org/project/reproject/}, keeping the original pixel size (in arcseconds), assuming a two-dimensional flux distribution (within a plane). To derive the orientation and size of the bar within the deprojected plane, we further assumed that the bar itself is represented by its main axis and length. In the following Sections, we will present barred galaxies aligning the deprojected main axis of the bar with the horizontal axis.

\subsection{Subsample definitions}
\label{sec:subsample}
\begin{figure}
\centering
\includegraphics[width=0.45\textwidth]{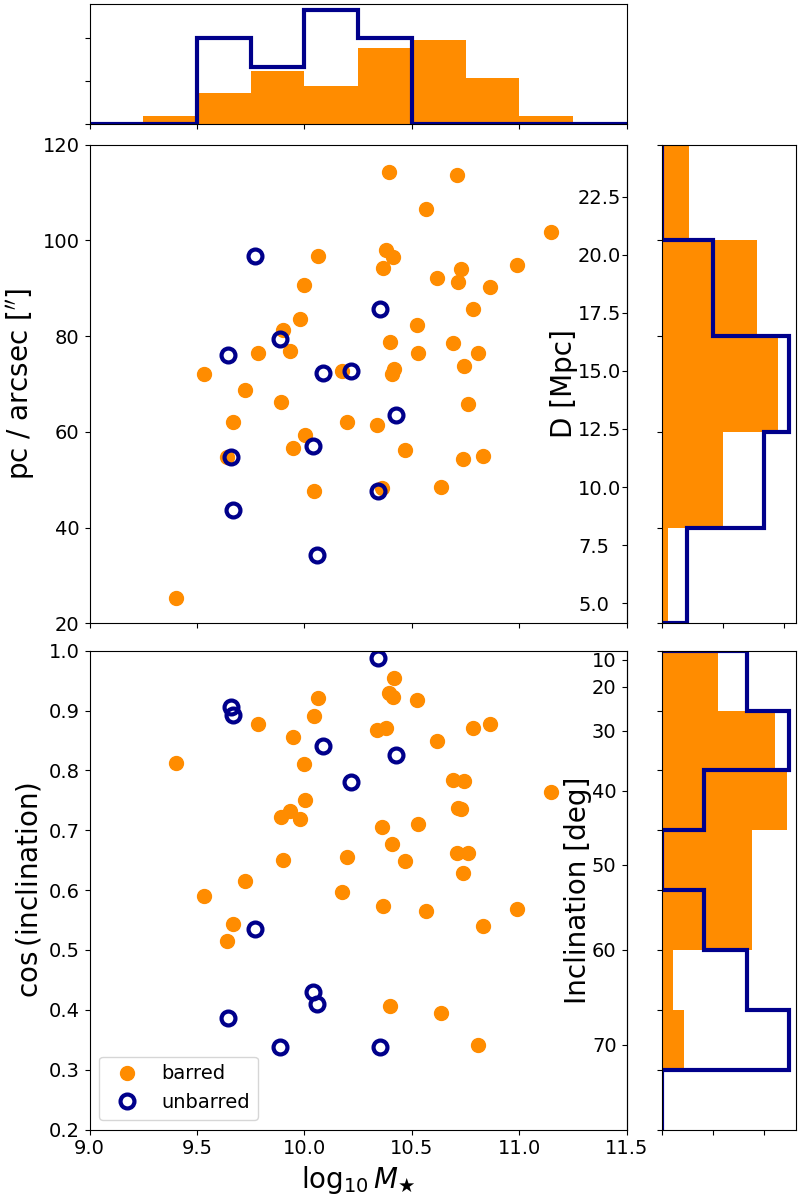}
\caption{Distribution of the 57 galaxies (45 barred, 12 unbarred) in the morphologically selected sample in terms of distance (top) and inclination (bottom) versus stellar mass. The distance is expressed here as a resolution in parsecs per arcsecond. The inclination is represented by its cosine, which serves as the stretching factor in image deprojection. Histograms on the right-hand side and top correspond to their respective axes and are normalised to the total counts in each category; note that unbarred galaxies are only 12, compared with 45 barred systems. Galaxies are coloured as barred (filled orange) versus unbarred (open blue).}
\label{fig:sample}
\end{figure}

We classified all 72 galaxies, extracting their physical \citep[stellar mass, following][]{Leroy+2021}, morphological (barred or unbarred) and intrinsic properties \citep[inclination, kinematic position angle; see][]{Querejeta+2021, Lang+2020}. We focused on star-forming main-sequence late-type galaxies, thereby removing 2 galaxies with negative Hubble stage $T$ values (considered "early-type"): NGC\,3626 and NGC\,4694. From this initial late-type sample of 70 galaxies, we further removed 9 galaxies showing clear signs of external disturbances and flagged them as peculiar: NGC\,1317, NGC\,1511, NGC\,2775, NGC\,3239, NGC\,3521, NGC\,4298, NGC\,4424, NGC\,4731 and NGC\,4826.
            
The presence of a bar in the remaining 61 galaxies, and their properties (bar length and position angles) were initially compiled from \cite{Querejeta+2021}, who heavily relied on measurements made by \cite{Herrera+2015} using the Spitzer Survey of Stellar Structures in Galaxies 3.6\,$\mu$m imaging \citep{Sheth+2010}. We then reviewed those measurements in the light of the newly acquired F300M JWST NIRCam imaging. Three barred galaxies that were not in the sample of \cite{Querejeta+2021} were added: NGC\,1808, NGC\,3344 and NGC\,3368. Two galaxies (NGC\,1385, and NGC\,5042) were re-classified as barred when examining the deprojected 3~$\mu$m images. Four galaxies, namely NGC\,1792, NGC\,2997, NGC\,3511 and NGC\,4951, were re-classified as uncertain. Of the four, only NGC\,3511 was classified as barred by \cite{Querejeta+2021}. Still, that bar is not readily visible in the 3~$\mu$m deprojected image, possibly because it is short and slightly elongated along the line of sight. This yielded a set of 45 barred galaxies and only 12 unbarred galaxies for consideration (confirmed bars in 79\% of the sample of 57 targets). This is the sample we will use for our study and in the following Sections of this paper. The bar radii $R_b$ and position angles have been extracted from \cite{Querejeta+2021} based on measurements by \cite{Herrera+2015}, except for four cases (NGC\,1385, NGC\,1808, NGC\,4569 and NGC\,5042) for which we obtained updated values after examination of the JWST 3\,$\mu m$ imaging.

In Fig.~\ref{fig:sample}, we show how the 57 galaxies we consider in the rest of the paper are distributed in terms of their stellar mass, inclination and intrinsic spatial resolution (given their individual distance). That sample peaks around a mass of $10^{10}$~\Msun, with reasonable coverage down to $10^{9.5}$~\Msun. It has only a handful of targets more massive than $10^{10.75}$~\Msun. We see relatively few with inclinations $< 10$\degr, a common issue that may be related, beyond the expected intrinsic rarity of such low inclination systems, to the difficulty of accurately measuring inclination for near face-on systems. The distribution of inclinations for the restricted set of barred galaxies is consistent with a flat distribution with a cut in inclination at about 70\degr, a criterion that defined the original PHANGS sample. However, almost half of the unbarred systems exhibit an inclination greater than 60\degr\ (or a cosine lower than 0.5): this may be driven by a bias in our ability to classify barred galaxies when they are more edge-on. It further implies that a fair fraction of the unbarred galaxies will have their imaging stretched significantly (by a factor of approximately 2 to 4) due to deprojection (see Sect.~\ref{sec:imaging}). Still, we have verified that this does not significantly impact our results (see Sect.~\ref{sec:limitations}). 

A more restrictive constraint is the absence of unbarred galaxies at stellar masses above 10$^{10.5}$~\Msun, which will prevent us from comparing barred and unbarred systems in the highest stellar mass bin. Such a lack of unbarred galaxies at high stellar mass may be only partly explained by the potentially higher fraction of barred systems in more massive discs \citep[see e.g.,][]{Sheth+2008, Melvin+2014} and the rarity of such (southern hemisphere) massive systems in the nearby universe \citep[e.g.,][]{Erwin2018}. We finally note that the spatial coverage of the galaxies in our sample varies (with typically 1 to 3 pointings with MIRI and 1 to 2 pointings with NIRCam). Only four targets, mostly on the massive side, have their bars not fully covered (NGC\,1097, NGC\,1365, NGC\,4569, NGC\,5248). Including or excluding those galaxies does not significantly impact our results.

\section{Results: the stellar mass trends}
\label{sec:res}

\begin{figure*}
\centering
\includegraphics[width=1\textwidth]{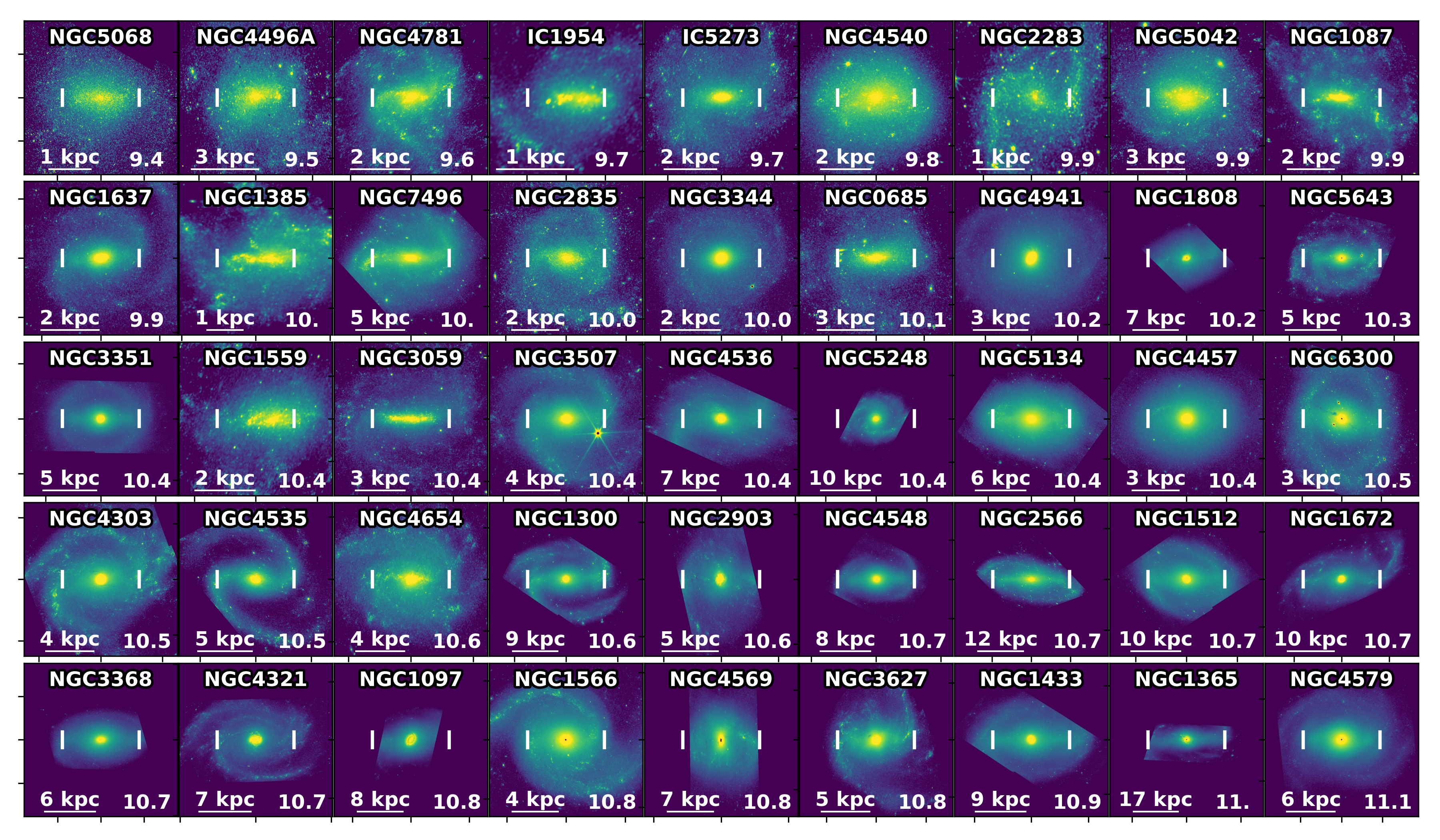}
\caption{Thumbnails of JWST F300M (3.0~$\mu$m) band deprojected images of the 45 barred galaxies considered in this paper. Galaxies are ordered from the top left to the bottom right by stellar mass M$_{\star}$: the value of $\log_{10}{({\rm M}_{\star})}$ is provided at the bottom right of each panel (rounded to the closest first decimal). The white bars indicate the radial extent of the bars in each panel; the field of view covers twice that radius. The scale in physical (kpc) units is provided at the bottom left of each panel.}
\label{fig:thumbnails_300}
\end{figure*}
\begin{figure*}
\centering
\includegraphics[width=1\textwidth]{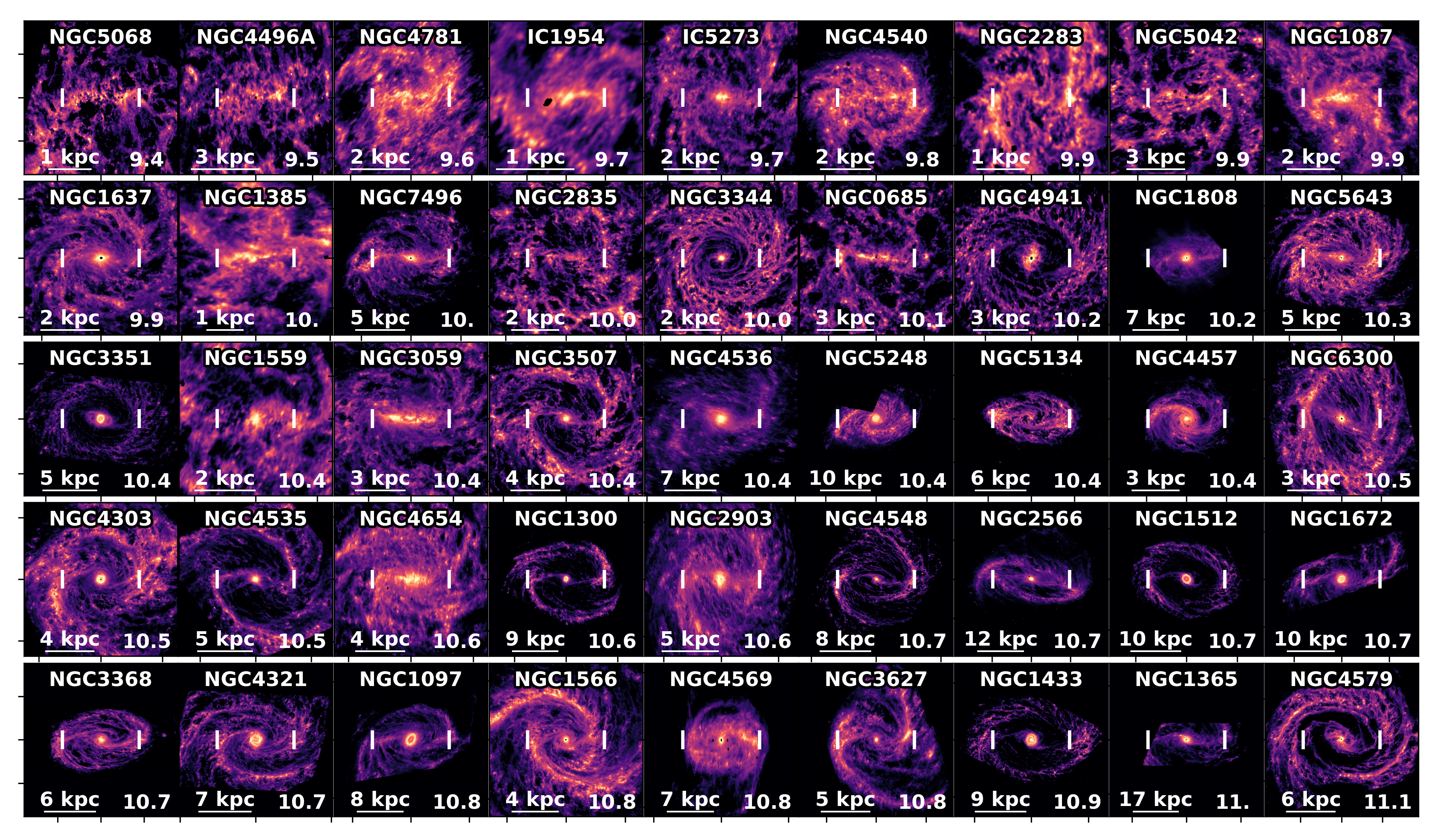}
\caption{Same as Fig.~\ref{fig:thumbnails_770s} but for the F770W$_{\rm ss}$ (starlight subtracted 7.7~$\mu$m) JWST band.}
\label{fig:thumbnails_770s}
\end{figure*}

In this Section, we review a few signatures of the stellar-mass-dependent change of regime in the bar-driven ISM distribution revealed by \citet{Verwilghen+25} via an extensive grid of star-forming main-sequence isolated disk hydrodynamical simulations covering a set of properties similar to the PHANGS sample of nearby galaxies. We first present all barred galaxies in our sample, using a stellar-mass ordering scheme, to briefly discuss their overall morphology. We then establish a systematic method for detecting the central structures in the processed 7.7~$\mu$m MIRI imaging, disentangling rings and inner discs. We finally derive the radial profiles and the density probability distribution function, consistent with expectations from numerical simulations \citep{Verwilghen+25}.

\subsection{Overall structuring}
\label{sec:struct}
In Figs.~\ref{fig:thumbnails_300} and \ref{fig:thumbnails_770s}, we present stellar-mass ordered JWST imaging in the F300M and F770W filters, after deprojection and, for the latter, continuum subtraction (see Sect.~\ref{sec:imaging}): all bars have been set up horizontally to ease the comparison, each square field with a size twice the estimated bar length (after deprojection).

Those figures first illustrate the apparent diversity of bar morphologies, as well as the shapes of stellar structures (at 3~$\mu$m) surrounding the bars, i.e., spirals, pseudo-rings, and clusters' density and distribution. Most importantly, we observe a change in the overall morphology with stellar mass, both in the proxy for the stellar continuum (F300M) and in PAH emission (F770W$_{\rm ss}$). 

Galaxies in our sample appear to fall into two general families, around the previously identified stellar mass threshold of about $10^{10}$~\Msun\ \citep{FraserMcKelvie+2020, Verwilghen+25}. Starting with the 3~$\mu$m imaging, bars in lower-mass systems tend to be elongated and thin (e.g., IC\,1954 in Fig.~\ref{fig:thumbnails_300}), often with a large family of superimposed clusters along the bar. At higher masses, bars appear generally smoother, often composed of an elongated and thick skeleton and a rounder central structure (e.g., NGC\,1300). The PAH emission in the bar regions of lower-mass galaxies is highly disordered, without much symmetry or a clear centre, while in higher-mass systems, the emission appears well-structured, more symmetric and often with spiral-like features, i.e. "bar lanes", extending from a bright central structure up to the bar ends. Four galaxies with stellar masses of $10^{10.1}$~\Msun\ and above (NGC\,0685, NGC\,1559, NGC\,3059 and NGC\,4654) stand out from those trends: all four have irregular and very elongated bar structures with a high level of PAH emission clustering within the bar, resembling the lower mass set: this will be discussed in Sect.~\ref{sec:disc}.

\subsection{Central reservoirs}
\label{sec:reservoir}
We now characterise the central structures in the PAH emission deprojected maps, adopting a simple fitting scheme to infer the existence (or absence) of inner density peaks or rings, and get an estimate for their extents, orientations and flattenings. 

We first extract a radial profile from the deprojected F770W$_{\rm ss}$ imaging using concentric circular rings. That profile is then used to first detect local peaks with a minimum prominence that can correspond to a ring-like structure (using the {\tt find\_peaks} routine from {\tt Scipy} package), as well as segments of roughly constant slope of the flux in log that may be associated with exponential disks.  Central peaks that are extended beyond the JWST Point Spread Function and bright (at least 1 dex) are tagged as inner "discs", while offset peaks corresponding to clear elliptic structures in the deprojected images are tagged as "rings". 

All detected rings are then characterised by selecting an annular region around each detected peak and using two-dimensional second-order moments (i.e., the inertia tensor). This provides a robust estimate of the position angle of the ring. We then proceed by computing a grid of ring-like masks with various axis ratios (100 bins between values of 0.1 and 1), always keeping the previously determined angle fixed. For each mask in that grid, we compute a score that follows the variance of 36 elliptical radii, derived by flux-weighted averaging of all pixels in 36 azimuthal bins. The axis ratio value of the mask with a minimum score is taken as the estimate of the axis ratio of the ring, as it represents the annulus that has the most constant elliptical radius as the azimuth varies. This scheme has similarities with the one developed by \cite{Lau+2012}, except that we only make use of the inertia tensor to obtain an estimate of the position angle of the ring. That choice was motivated by the fact that second-order moments can lead to rather significantly varying values of axis ratios when using an iterative scheme.

All detected inner discs are fitted using a double exponential profile, including one inner exponential associated with the inner discs and one outer exponential associated with the large-scale disc or background.
Examples of detected radial structures are provided in the Appendix in Fig.~\ref{fig:radialfit}. The results from the radial structure detection process are further illustrated in the Appendix, in Fig.~\ref{fig:rings770s} and \ref{fig:rings300} and tabulated in Table~\ref{tab:sample}. 

Five galaxies among the 45 barred systems could not be properly fitted due to a large number of saturated pixels in their central regions: NGC\,1637, NGC\,4569, NGC\,4941, NGC\,6300, and NGC\,7496, potentially partly due to the presence of a central active galactic nucleus (AGN). 15 of the 45 barred galaxies have no detected central structure (an extended disc or a ring), with 10 of these having stellar masses below or equal to $10^{10}$~\Msun. Out of the 12 galaxies that have a stellar mass below $10^{10}$~\Msun, two of them (NGC\,1637 and NGC\,7496) have a saturated centre, preventing the detection algorithm from being applied, and the remaining ten show no evidence for a central detected structure. Only four galaxies that have stellar masses significantly above that threshold, namely NGC\,0685, NGC\,1559, NGC\,3059 and NGC\,4654, have no specific central structures. Those four hosts are the ones already mentioned in Sect.~\ref{sec:struct} that have very elongated and irregular bars with a high level of clustering at 3~$\mu$m: in other words, the absence of a central extended source or ring and the overall structuring of their bars resemble the lower mass galaxy set in our sample.
Conversely, only galaxies with stellar masses above or equal to $10^{10}$~\Msun\ have a detected central bright disc-like or ring-like structure (the lowest mass one in our sample being NGC\,3344 at $10^{10.1}$~\Msun).

The ellipticity of detected rings is generally very small with typical values around 0.05, and a maximum at $\sim 0.2$ for NGC\,1097. The relative orientations of the detected rings with respect to the bar major-axis appear to be randomly distributed between 0 and 180\degree. However, this should not come as a surprise, given the low ellipticities we derived.

\subsection{Radial redistribution}
\label{sec:radial}

\begin{figure}
\centering
\includegraphics[width=0.5\textwidth]{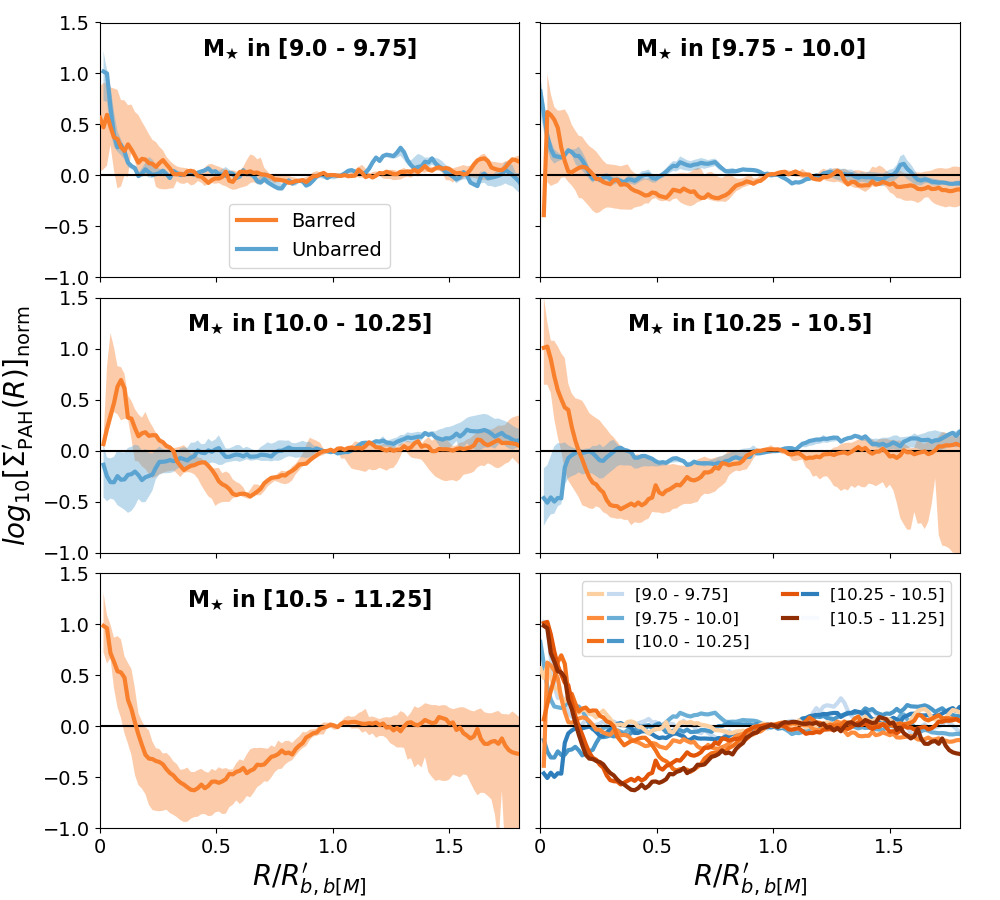}
\caption{Radial PAH emission (F770W$_{\rm ss}$ band) profiles of barred (orange lines) and unbarred (blue lines) galaxies: each line corresponds to the median of all barred galaxies within a stellar mass bin (as indicated by the legend in each panel). In the bottom right panel, all lines are shown together to illustrate the trend. Each of the other five panels shows a single stellar mass bin for both the barred and unbarred samples. In those panels, the transparent-filled regions correspond to the first and third quartiles (25 and 75\% percentiles) around the median. All individual galaxy profiles have been divided by an exponential profile before the averaging using radial scale length fitted on the F300M band, and then normalised by the value at $R = R_b^{\prime}$: for unbarred galaxies, $R_{b[M]}^{\prime}$ as a proxy \citep[see text, and][]{Erwin2019}. Note the absence of unbarred galaxies in the most massive stellar mass bin.}
\label{fig:radial_bars}
\end{figure}

We now turn to the radial profiles derived from azimuthally averaged, deprojected PAH emission images. For each galaxy in the sample of 45 barred and 12 unbarred galaxies, we compute the average radially binned surface brightness profiles. For barred galaxies, those profiles are then normalised at $R_{b}^{\prime}$, the deprojected bar radius for all barred galaxies, thus emphasising the bar region we are interested in.

For unbarred galaxies, we derived a proxy radius ($R_{b[M]}$) using a mass-based scaling law fitted to our barred sample. This approach, adapted from Erwin (2019), employs a broken power-law with a transition mass ($M_{brk}$) to assign a characteristic 'bar-equivalent' scale to unbarred discs. Using the bar radii from our barred galaxy sample, we find $\log_{10}(R_{b[M]}^{\prime}) \sim \alpha + \beta \cdot \log_{10}(M_{\odot})$ with $\alpha = -4.8$ and $\beta = 0.5$ for stellar masses below $M_{brk} = 10.16$, and $\beta = 0.8$ otherwise \citep[see][for details]{Erwin2018, Erwin2019}. That scaling relation can then be applied to unbarred galaxies, using their respective stellar mass estimates. While this method is the most direct, we have tested other proxies that rely on scaling relations involving the disc scale length $R_d$ or on the deprojected radial F300M profiles, and all yielded consistent results.

Fig.~\ref{fig:radial_bars} shows the result of averaging the radial profiles for all galaxies within four bins of stellar mass, separating the barred systems (top panel) and the unbarred systems (bottom panel). Following the approach of \cite{Verwilghen+25}, we normalised all profiles by an exponential law where the scale length $R_d^{\prime}$ was fitted onto the F300M band radial profile: this serves to emphasise departures from that reference profile. We first need to flag again the significantly smaller number of unbarred galaxies in each mass bin (typically 2 to 4), and the unfortunate absence of systems in the largest mass bin. Nevertheless, we observe a general trend for massive barred galaxies to exhibit a region of lower PAH emission in the radial range [0.2 -- 0.8]\,$\cdot\ R_b^{\prime}$, accompanied by a central density increase within the inner $0.2 \cdot R_b^{\prime}$. The central peak and depression (relative to the reference exponential) have larger amplitudes for the largest stellar mass bin, respectively reaching $\sim +1$ and $\sim -0.6$~dex. Beyond $R = R_{b}^{\prime}$, the normalised profiles are consistent with a constant (within $1\,\sigma$). This is in stark contrast with the unbarred sample, which exhibits profiles mostly consistent (within $1\, \sigma$) with an averaged constant normalised PAH emission density. Only within the central $0.15 \cdot R_{b[M]}^{\prime}$, the radial PAH profiles of unbarred systems show either a peak (for the two lowest mass bins) or a depression (for the two largest mass bins). It is worth noting that all four galaxies that show a central extended hole in their PAH distribution are unbarred: NGC\,628, NGC\,2090, NGC\,2775, and NGC\,3521; the last two are flagged as peculiar. Hence, the azimuthally averaged, deprojected, PAH profiles for unbarred galaxies are, on average, consistent with a simple exponential over a large fraction of the disc. It is also worth noting that both the unbarred and barred galaxies in the lowest stellar mass bin ($\log_{10}({\rm M}_{\star} / {\rm M}_{\odot}) < 9.75$) look very similar, including for the central peak within about 15\% of $R_{b}^{\prime}$ (or $R_{b[M]}^{\prime}$). 

\subsection{The surface density PDF}
\label{sec:pdf}

\begin{figure}
\centering
\includegraphics[width=0.5\textwidth]{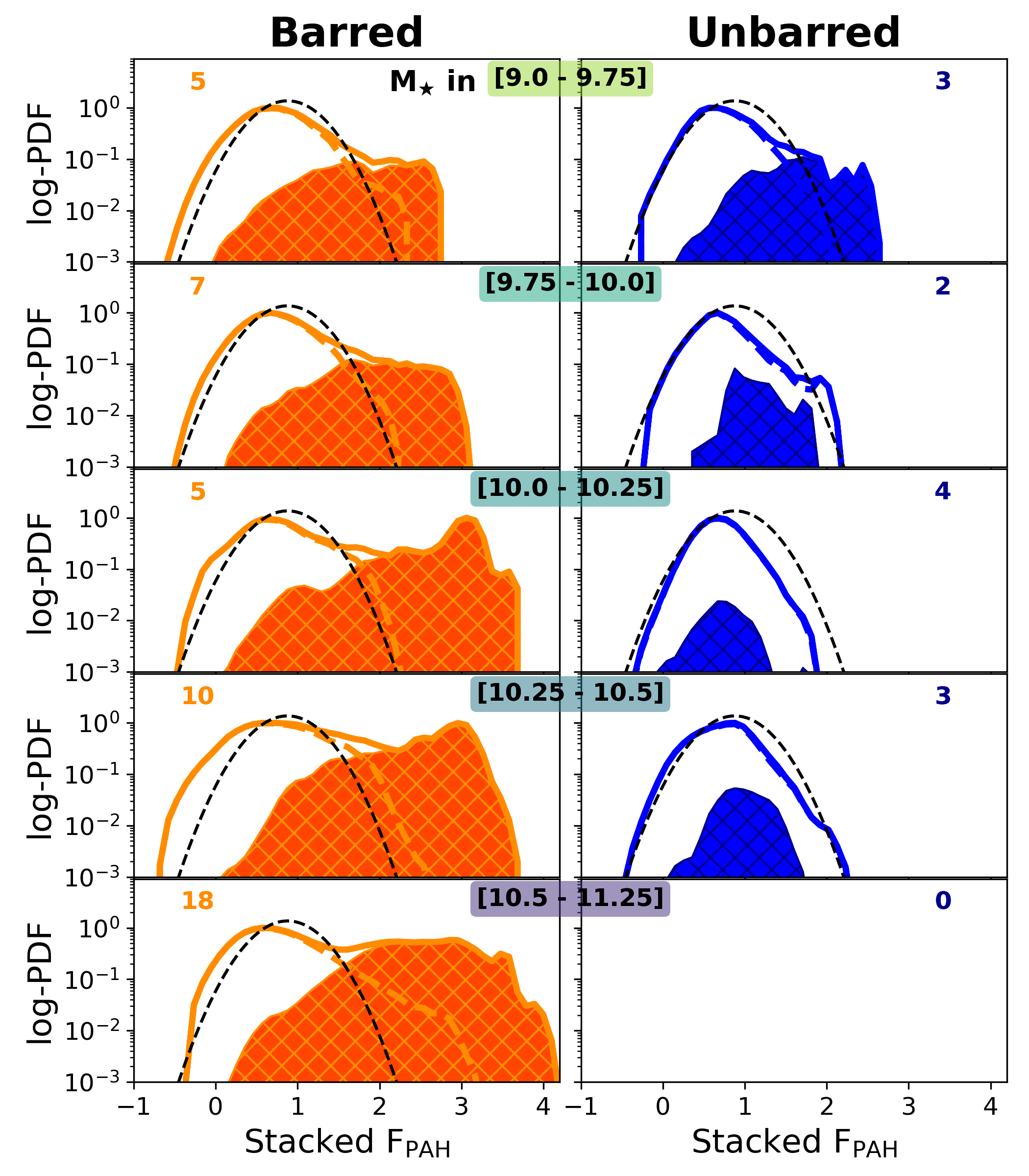}
\caption{Stacked and shifted PAH emission density PDFs using the F770W$_{\rm ss}$ JWST band in bins of mass, going from the least massive (top) to the most massive (bottom) targets, with the stellar mass bin indicated in the middle of the two panels of each row. The left (right) panels correspond to the barred (unbarred) subsample. The value at the top left (right) of each panel is the number of targets stacked for the barred (unbarred) systems. The thick coloured lines correspond to the PDF for all pixels within $R_{b}^{\prime}$ for barred galaxies, and within $R_{b[M]}^{\prime}$ for unbarred galaxies. The filled hatched (orange or blue) areas are PDFs restricted to a radius of 15\% of the respective reference radii. A fixed log-normal function peaking at an abscissa of 1 (with which individual profiles have been aligned) is plotted as a black dashed line in each panel for reference.}
\label{fig:PDF}
\end{figure}

In Fig.~\ref{fig:PDF}, we present the density probability distribution function (PDF), namely the histogram of surface densities, using the deprojected F770W$_{\rm ss}$ imaging. As in Fig.~\ref{fig:radial_bars}, we split the sample into four stellar mass bins, and compute both the averaged PDF (thick lines) within a circular area defined by $R = R_{b}^{\prime}$ ($R = R_{b[M]}^{\prime}$) for the barred (unbarred) galaxies, as well as from the corresponding inner 15\% of that region (hatched area). Before averaging the PDFs in each mass bin, all individual PDFs have been renormalised (shifted on a log scale) using the log-normal emission peak fitted to the lower-density part of the PDF: this is a necessary step to remove the dependence on varying gas fractions and total masses, which would otherwise yield artificial multiple-peaked PDFs when stacked.

Unbarred systems have density PDFs that are close to log-normal, with only a small tail towards higher densities in the lowest mass bin. The barred galaxy sample systematically exhibits a significant flattening of its averaged PDF or even a secondary peak at high densities. The amplitude and density of this plateau or peak increase with stellar mass; in the highest mass bin, the PDF is flat over two orders of magnitude. Remarkably, most, if not all, of the high-density PAH emission in the barred galaxies corresponds to the central 15\% of the bar region (hatched areas). In contrast, for the unbarred systems, the central region is consistent with a log-normal distribution in the two intermediate-mass bins. As for the radial profiles, barred and unbarred galaxies in the lower mass bin exhibit very similar PDFs.

In other words, it is difficult or even impossible to distinguish a barred from an unbarred system based solely on the tracers associated with the F770W$_{\rm ss}$ radial profile or PDF for masses below $10^{9.75}$~\Msun. Such a separation is obvious for larger mass systems. In the following, we will label the regime acting on low stellar mass galaxies, the low-mass, feedback-dominated regime.

\section{Discussion and Conclusions}
\label{sec:disc}

\subsection{The nature of the F770W$_{\rm ss}$ emission}
\label{sec:f77ss}
In this Section, we assess whether the structural and radial trends identified in the previous sections are robust or if they are significantly biased by our choice of tracer.

The $F770W$ filter includes, in addition to emission from PAHs, contributions from both the underlying stellar population and small dust grains. Our data processing applies a rigorous starlight subtraction to yield $F770W_{\rm ss}$ deprojected images, meaning the underlying old stellar continuum is already accounted for. However, the resulting $F770W_{\rm ss}$ emission primarily traces the $7.7\,\mu\text{m}$ Polycyclic Aromatic Hydrocarbon (PAH) complex. Because PAHs are stochastically heated by the interstellar radiation field (ISRF), the observed emission reflects a degenerate interplay between three distinct physical properties \citep[][L. Hands et al., in prep.]{Donnelly+2025, Chown+2025b} : 
(i) PAH column density \citep[which typically scales with total gas column density;][]{Leroy+2023, Sandstrom+2023a}; 
(ii) the intensity and hardness of the ISRF \citep[which dictates dust heating and relative band strengths; e.g.,][]{Draine+2021} and (
iii) the PAH ionisation state (which strongly modulates the $7.7\,\mu\text{m}$ feature). Because these properties vary across galactic environments, the translation from $F770W_{\rm ss}$ intensity to absolute gas mass is not strictly uniform. Below, we evaluate how these environmental variations impact our interpretation in two extreme regimes: galaxy centres and "star formation deserts."

In the central molecular zones (CMZs) of massive barred spirals, the ISRF can be highly intense, up to $\sim 10^2\times$ the Solar neighbourhood value \citep{Pathak+2024, Pathak+2026}. Such intense radiation fields naturally boost the PAH emission per unit mass, potentially overestimating the gas concentration. However, empirical work by \citet{Chown+2025a} demonstrates an excellent correlation between CO emission (tracing molecular gas mass) and $F770W_{\rm ss}$ in galaxy centres, albeit with an offset normalisation compared to galactic disks and a $\sim 0.2$ dex scatter linked to the specific SFR. Physically, the intense star formation required to produce these extreme radiation fields cannot exist without a corresponding reservoir of high-density gas. Therefore, the central enhancements we observe in $F770W_{\rm ss}$ represent a coherent, mutually reinforcing response: the bar concentrates the gas, which triggers star formation, which in turn elevates the ISRF. Both effects act in tandem to illuminate the structural concentration of material.

Conversely, in "star formation deserts" found along bars, the $F770W_{\rm ss}$ emission is often noticeably suppressed. Recent work suggests this suppression may be partly an ionisation effect rather than a pure gas deficit. Spectroscopic tracers show that the $F770W/F1130W$ ratio correlates with the hardness of the radiation field \citep{Baron+2024, Baron+2025}. Because the $11.3\,\mu\text{m}$ and $3.3\,\mu\text{m}$ features arise from more neutral PAHs, a softer radiation field (dominated by older stellar populations) shifts the ionisation state, suppressing the $7.7\,\mu\text{m}$ ionised feature relative to other bands \citep{Baron+2025, Pathak+2026, Koziol+2026}. Consequently, the lack of young stars in these regions artificially deepens the observed "deserts" in $F770W_{\rm ss}$ images. However, according to the results by \cite{Baron+2025}, anomalous PAH line ratios are not observed in most of the massive barred galaxies in the sample, and when present are localised close to or along the gas bar lanes.

Because of these localised variations in heating efficiency and ionisation, we intentionally maintain an observational framing for our results rather than converting $F770W_{\rm ss}$ directly into a physical gas mass ($\Sigma_{\rm gas}$). Crucially, however, these environmental factors amplify rather than cancel the underlying gas distribution signals. The structural features we measure in $F770W_{\rm ss}$ map onto real, physically driven concentrations and deficits of material orchestrated by the galactic bar, rather than being artefacts of competing, coincidental physical effects.

\subsection{Limitations}
\label{sec:limitations}
Results presented in the previous Section emphasise the change in the mass-dependent structuring of the ISM as traced by PAH emission driven by bars. The main limitations of this work lie in the relatively small and sparse sample of galaxies along the star-formation sequence, as well as in uncertainties associated with image deprojection. 

The sample of unbarred galaxies is only a third in number of the sample of barred galaxies. While this is consistent with expectations from the fraction of barred galaxies in the nearby universe \citep{Sheth+2008, Erwin2018}, it still means we use only a handful of unbarred targets in each mass bin and none in the highest mass bin. We do not think that we have missed unbarred systems (false positives), but we may have missed a few barred ones (false negatives). Even though we based our classification on both the 3~$\mu$m and 7.7~$\mu$m JWST images, there could be a bias coming from circular reasoning: galaxies with disordered PAH within the central few kiloparsecs may more often be classified as unbarred. We find that all barred galaxies with stellar masses below $10^{10}$~\Msun\ have an unstructured central region, as traced by PAH. Hence, the question we need to address is: are the unbarred, higher-mass galaxies misclassified? Scrutinising the 3~$\mu$m JWST images, we can rule out the presence of large, strong bars. Still, we cannot discard the possibility that some of the classified unbarred galaxies, like NGC\,1546, NGC\,2090 or NGC\,4254, have weak bars with radii significantly smaller than 1~kpc. While those would depart from the bulk population of nearby barred galaxies \citep{Erwin2024}, it is important to keep this fact in mind when flagging galaxies as ``unbarred''.

The simplified deprojection scheme that we apply (including the assumption of a two-dimensional flux distribution) introduces potential uncertainties. 
A more general approach would have been to represent the bar as a three-dimensional ellipsoid. However, this would imply assessing its thickness and height, a degenerate problem, and would not have a significant impact on our results. 
Our assumptions also affect the intrinsic orientation of the bar within the deprojected plane, but this does not significantly affect the results in Sections~\ref{sec:radial} and \ref{sec:pdf}, as the analysis is not azimuth-dependent. When focusing on galaxies with inclinations below 45\degr\ ($\cos{(\rm{inclination})}$ above $\sim 0.7$, see Fig.~\ref{fig:sample}), we do not detect any significant differences in the results reported in the previous sections. Trends seen in the radial PAH emission profiles (Fig.~\ref{fig:radial_bars}) are conserved. We are thus confident that, while a few individual cases may be slightly impacted, our results are robust against deprojection assumptions.

\subsection{The dips, or bar star-formation deserts}
\label{sec:deserts}
In Section~\ref{sec:radial}, we have shown that the radial distributions of PAH emission in our barred galaxy sample exhibit a relative depression corresponding to the reduction in emission within the bar region and away from the main axis of the bar. The relative amplitude of the dip appears as a continuously increasing function of stellar mass: it is significant for hosts with a stellar mass above $10^{10}$~\Msun, and can reach up to 0.5~dex on average within the highest stellar mass bin. We also see a mass-dependent trend on the central inner PAH peak. Those trends are not observed in the unbarred sample.

Such a behaviour is reminiscent of what has been identified by, e.g., \citet{DiazGarcia+2020}, in the far and near UV light distribution for a sample of hundreds of galaxies: the deficit of UV light within the inner few-kpc region is not observed in the unbarred systems (see e.g., their Fig.~6), or in lower mass galaxies (see also their Fig.~5). It is also consistent with the picture emerging from the observed distribution of ionised gas by \cite{FraserMcKelvie+2020}, which focuses on the centre and the bar ends, for galaxies more massive than $10^{10}$~\Msun, albeit at a significantly lower spatial resolution ($\gtrsim 1 kpc$).

Such dips have often been connected with the so-called ``star formation deserts'' (SFDs) in bars \citep{James2018}, a circum-bar region away from the major-axis of the bar where new stars and gas are scarce. Such SFDs have been studied via observations \citep[e.g.,][]{James2018, DiazGarcia+2020, Neumann+2020, Kim+2025, Krishnarao+2022} and simulations \citep[e.g.,][]{Donohoe-keyes+2019}, and flagged as the result of the gravitational (dynamical) influence of the bar. This was initially emphasised via the H$\alpha$ radial profiles by \cite{James+2009}, who noticed that barred galaxies classified as SBb and SBbc in their sample had a profoundly different profile than the unbarred hosts. 

Those results emphasise that the distribution of star formation tracers is significantly depleted within the bar region away from the bar major-axis (within the SFD), relative to along the bar itself for relatively massive hosts. In this study, we expand on this using high-resolution PAH imaging and further demonstrate a strong mass trend in a barred galaxy sample. This provides a direct observational reminder that bars in high stellar mass systems can be efficient actors in the redistribution of gas, dust and subsequent star formation \citep[e.g.,][]{Roberts+1979, Sanders+1980, Athanassoula1992, Kuno+2007} and in the creation of central stellar rings and discs \citep[][and references therein]{Schultheis+2025}. Such structures are observed in various tracers, including e.g., UV \citep{Comeron+2010} or molecular gas \citep{Stuber+2023, Gleis+2026} and can contribute significantly to the molecular gas content and star formation rates of their host galaxies \citep{Querejeta+2021, Schinnerer+Leroy2024, Gleis+2026}.

\subsection{PDFs and star formation regimes}
\label{discPDF}
In the previous Sections, we suggested that the radial PAH emission distribution of low-mass galaxies (with masses below $\sim 10^{10}$~\Msun) appears independent of whether the host galaxy is barred. This is also visible, as emphasised in Fig.~\ref{fig:PDF}, via the observed characteristics of the surface density PDF within the bar radius. Unbarred systems and low-mass barred galaxies maintain a near-log-normal PDF, whereas massive barred galaxies exhibit a significant departure from this distribution. The observed flattening and secondary high-emission peaks in the PAH emission PDFs for massive barred galaxies (Fig.~\ref{fig:PDF}) therefore provide a quantitative signature of this regime change (in the lower-mass star formation regime).

This echoes the work of \cite{Pathak+2024}, which shows that PDFs, traced via mid-infrared emission across four JWST bands, are generally composed of two distinct but overlapping components: a log-normal diffuse component and a power-law high-density tail. At 21~$\mu$m, the power-law tail is mostly associated with star formation, as emphasised by \cite{Pathak+2024}. At 7.7~$\mu$m, we more closely follow the PAH distribution, and the lower-density log-normal component becomes more dominant. In our analysis, we see (Fig.~\ref{fig:PDF}) that most, if not all, of the high-density PAH regions lie within the central 15\% of the galaxy reference radius (the bar radius when the galaxy is barred; see Section~\ref{sec:pdf}). This result is consistent and significantly expands on the work by \cite[][see their Sect.~3.6]{Pathak+2024}, who studied the first 19 targets of the PHANGS-JWST Cycle~1 Treasury sample.

Like \cite{Pathak+2024}, we do not detect a significant change in the log-normal component of the PDF with stellar mass, both in barred and unbarred galaxies of our sample. In unbarred galaxies, the inner and outer PDF seem to be indistinguishable (Fig.~\ref{fig:PDF}), a result that is also illustrated at low mass with the radial profiles presented in Fig.~\ref{fig:radial_bars}. While we obviously do not expect any typical bar-driven structures (e.g., rings, dust lanes) in unbarred galaxies, we now suggest that such signatures in low-mass galaxies, either barred or unbarred, are basically weak or absent. A more qualitative scrutiny of the deprojected maps presented in Fig.~\ref{fig:thumbnails_770s} and in Appendix~\ref{sec:appthumb} confirms that bars are not conspicuous in the PAH map of low-mass galaxies. 

This is consistent with the results from \cite{Stuber+2023} and \cite{Gleis+2026} who find bar lanes and rings as traced by molecular gas almost exclusively in massive galaxies (with stellar masses above $10^{10}$~\Msun). It seems in line with trends observed in the molecular gas density PDFs as emphasised by \cite{Sun+2018}. It also provides a new perspective on the discrepancy mentioned by \cite{Erwin2018} who noted that bars are common in low-mass galaxies, with a maximum bar frequency around a stellar mass of 9.7 (in $\log_{10}({\rm M}_{\star} / {\rm M}_\odot)$) while other surveys reported that the bar frequency increased above $\sim 10^{10}$~\Msun: bars at low stellar mass are shorter but also, most importantly, much harder to trace using the distribution of dust, gas or star formation. 

\subsection{Mass-dependent regimes of bar-driven ISM evolution}
\label{speak}
Using a grid of hydrodynamical simulations conducted with the adaptive mesh refinement code RAMSES \citep{Teyssier2002}, \cite{Verwilghen+25} studied the evolution of star-forming main-sequence discs, following global properties (size, stellar mass, gas fraction) of galaxies in the PHANGS sample. Most of those simulations develop a bar in a few hundred million years, allowing the probing of bar-driven secular processes \citep[see further details in][]{Verwilghen+25}. The continuous but steep transition witnessed in the previous Sections using JWST imaging is consistent with the findings of \cite{Verwilghen+25}, where the emergence of high-density gas reservoirs is primarily localised within the central 15\% of the bar radius. Figs~\ref{fig:radial_bars} and \ref{fig:PDF} are further reminiscent of Figs.~3 and 5 of \cite{Verwilghen+25}, who traced the radial profiles and PDFs of mass-weighted gas densities in their set of generic isolated disc simulations. \cite{Verwilghen+25} demonstrated that, in the simulations, the physical origin of the change (around a stellar mass of $10^{10}$~\Msun) in the ISM distribution depends on the relative contributions of gravity and stellar feedback.

Our results reinforce the hypothesis that a stellar mass of about $10^{10}$~\Msun\ acts as a "watershed" for barred galaxy evolution. Below this threshold, the PAH emission remains clumpy and disorganised, with little to no detectable central peaks. In these low-mass systems, stellar feedback appears potent enough to disrupt the gas flows organised by the stellar bar, effectively suppressing the observable imprint of the bar's presence in ISM tracers. This is the low-mass feedback-dominated regime, which yields radial profiles and PDFs nearly indistinguishable from those of unbarred galaxies. In that regime, gas properties such as the PDF (see Sect.~\ref{sec:pdf}), as well as, e.g., the virial parameter $\alpha_{vir}$ or the Mach Number $\cal M$, are similar within the bar, outside the bar, and, most remarkably, in the inner regions before the formation of the bar itself \citep[see e.g.,][]{Verwilghen+25, bland-hawthorn_turbulent_2025}.

Conversely, as galaxies cross the $10^{10}$~\Msun\ threshold, the increase in stellar mass deepens the potential well to the point where bar-driven torques dominate over feedback. This leads to the formation of a central mass concentration and thus to orbit families that allow the build-up of inner rings or discs \citep[as seen in, e.g., Fig.~4 of][]{Verwilghen+25}. The bar's presence is revealed by highly structured ISM features, symmetric spirals, and prominent central gas reservoirs and stellar structures (discs or rings).

The presence of "outliers" like NGC\,1559 and NGC\,4654, massive galaxies that nonetheless exhibit morphologies more typical of low-mass barred systems, could be explained in several ways. One option is that their stellar masses are overestimated due to an overestimated distance, something that would require a new analysis of distance indicators. A second option is that, since stellar mass drives the timescale associated with the development of central bar-driven structures \citep{Verwilghen+25}, the bars in those "outliers" are dynamically young. This may relate to the relatively high specific star formation rate (sSFR) for, e.g., for NGC\,1559, NGC\,3059 and NGC\,4654. However, we should note that other targets with similar or higher sSFR (e.g., NGC\,1365 or NGC\,1672) have the typical PAH emission central structure of massive galaxies. This emphasises the importance of considering the internal processes associated with the dynamical secular (isolated) evolution of galactic discs together with external processes that may lead to late bar formation, such as accretion or interactions. 

The resulting low-redshift galaxy population is the result of a potentially non-monotonic, complex assembly and formation history. Hence, the stellar-mass-dependent timescales associated with the ISM redistribution emphasised by the work of \cite{Verwilghen+25} studying ideal isolated disc galaxies should be considered together with the occurrences and timescales attached to such external perturbing events \citep[e.g.,][]{BournaudCombes2002, Fragkoudi+2025, Lu+2025}. A third option is that the existence of such outliers could imply that while stellar mass is the primary driver, local conditions or other environmental properties may temporarily sustain the feedback-dominated regime. This is also a reminder that the transition at about $10^{10}$~\Msun\ is expected to be continuous, even if apparently steep, one driven by the relative balance between various physical processes.

\subsection{Timescales and the masking of bar signatures}
The detection of central discs and rings exclusively in galaxies above the mass threshold points to a rapid phase of secular evolution. Numerical simulations suggest these structures start their build-up on timescales of a few bar rotations (or $\sim$0.5–1 Gyr) in more massive hosts \citep[see, e.g.,][]{Verwilghen+24}. It takes at least 4 times longer (both in absolute terms or relatively to the orbital time or bar rotation period) for systems below the $10^{10}$~\Msun\ stellar mass threshold \citep{verwilghen_simulating_2025}. The bright, compact nature of these central reservoirs (Sect.~\ref{sec:radial}) suggests they are the primary sites for the bursty build-up of secondary stellar components, potentially leading to the formation of central stellar discs or rings \citep[see e.g.,][; and Neumann et al. in prep.]{Bittner+2020}.

The lack of such structures in low-mass barred galaxies implies that secular evolution induced by the bar proceeds at a significantly slower pace, often exceeding many Gyr. This further means that for a significant fraction of their lifetimes, low-mass barred galaxies exist in a "proto-barred" state when just considering their ISM state, where the bar's influence is present as a stellar (gravitational potential) skeleton but yet to be significantly manifested in the radial gas redistribution and the subsequent evolution of the central stellar structures.

It is worth noting the parallel between the feedback-dominated regime and predictions from the analytic model developed by \cite{hayward_how_2017} for the driving of outflows by stellar feedback: the model predicts a regime change that depends on the gas fraction and stellar mass. At low redshift, the model predicts a stellar mass threshold of $\sim 10^{10}$~\Msun\ above which outflows are efficiently suppressed. In other words, we expect lower-mass galaxies on the star-formation main sequence to have more efficient outflows driven by stellar feedback. This is further emphasised by the fact that galaxies below a stellar mass of $\sim 10^{10}$~\Msun\ exhibit, on average, an increasing gas fraction \citep[see e.g., Fig.~1 of ][]{Verwilghen+24}. Beyond the potential connection with the masking of bars' signatures, the processes mentioned by \cite{hayward_how_2017} may drive the trend for other galactic properties such as the metal content, as proposed by, e.g., \cite[][; their Sect.~5.4]{Blanc+2019}. In such a context, the stellar mass threshold is expected to shift to higher values at higher redshifts as the gas fraction in the population of star-formation main-sequence galaxies increases.

The mass-dependent timescale transition described here has significant implications for our interpretation of galaxy evolution across cosmic time. If bars in lower-mass or gas-rich systems are difficult to detect via traditional ISM tracers or star-formation maps, we may be consistently underestimating the bar fraction in the early universe or in low-mass dwarf populations.

As we probe the morphology of galaxies at higher redshifts \citep[see e.g.,][]{Menendez-Delmestre+2007,LeConte+2024}, the distinction between a truly unbarred galaxy and one in which the bar is "invisible" to the gas becomes crucial. Our findings suggest that a lack of central gas concentration or dust or gas "lanes" is not necessarily evidence for the absence of a bar. Instead, it may reflect a regime where the interplay between gravity and feedback has not yet allowed the bar to assert its dominance over the ISM. Understanding this transition is key to addressing the varying timescales associated with bar-driven processes and to mapping the true history of dynamical instability in disc galaxies. Simulations examining the evolution of discs in regimes that better reflect the high-redshift universe (high gas fraction, shorter dynamical timescales, disturbed assembly and accretion histories) and scrutinising the dependence on the underlying gravitational potential would be key to better understanding the evolution of discs and the impact of bars.

\section{Conclusions}
\label{sec:conc}
In this paper, we have leveraged high-resolution JWST imaging from the PHANGS-JWST survey to characterise the ISM distribution in a sample of 57 nearby disc galaxies (out of 72). By focusing on PAH emission as a proxy for the cold ISM, we have re-examined the impact of stellar bars identified in the near-infrared across a wide range of stellar masses ([9.5 -- 11.5] in $\log_{10}({\rm M}_{\star} / {\rm M}_{\odot})$

Our main conclusions are as follows:
\begin{itemize}
    \item The $10^{10}$~\Msun\ watershed: We confirm that a stellar mass of $\sim 10^{10}$~\Msun\ marks a continuous but abrupt transition in how bars organise the ISM. Above this threshold, barred galaxies exhibit well-structured, symmetric PAH distributions with prominent central reservoirs (e.g., discs and rings). Below this mass, the ISM distribution remains clumpy and unstructured, appearing largely agnostic to the presence of a stellar bar.
    \item The suppression of bar-driven signatures: For barred galaxies with stellar masses below $\sim 10^{10}$~\Msun, the radial PAH emission profiles and density PDFs are nearly indistinguishable from those of unbarred systems. This likely results from the dominance of stellar feedback over gravity in lower-mass potential wells, preventing the bar from establishing the classic flow patterns seen in more massive hosts.
    \item The emergence of bar deserts: In massive barred galaxies, we observe a systematic "dip" or depletion of PAH surface density in the radial range of [0.2–0.8]~R$_{b}$. These "bar deserts" increase in amplitude with stellar mass, reflecting the efficient redistribution of cold gas by the bar potential into central reservoirs within a radius of 15\% of the bar radius $R_b$.
    \item Central engines and secular evolution: The presence of bright central discs and rings is almost exclusively restricted to galaxies above the mass threshold. As suggested by simulations, the formation of these structures on relatively short timescales ($\sim 0.5-1$~Gyr) in massive galaxies indicates a rapid phase of secular evolution that is significantly delayed or suppressed in lower-mass systems.
    \item Anomalous massive bars: We identify a small subset of massive galaxies (e.g., NGC\,1559, NGC\,3059, and NGC\,4654) that host irregular, clumpy bars despite their high stellar mass. Assuming their distance (hence their stellar mass) estimates are correct, these objects suggest that while mass is the primary driver of the emergence of structured bar-driven features, secondary factors such as environment or gas fraction can allow the feedback-dominated regime to persist.
\end{itemize}

Our results provide a relevant framework for understanding bar-driven evolution. The fact that bars can remain less conspicuous in ISM tracers at low stellar masses suggests that the bar fraction and their impact on galaxy growth may be systematically underestimated in high-redshift or low-mass galaxy populations. It also provides a new, mass-dependent perspective on the secular evolution of discs under the influence of bars. As bars are active drivers of the redistribution of the ISM and thus of the subsequent star-forming regions, cosmological simulations must therefore resolve such (mass-dependent) processes when they have the ambition to describe and reflect the evolutionary history of galaxies properly. This is a challenge for existing projects \citep{Zhao+2020, Lopez+2024, ansar_bar_2025}, as it may relate to spatial and mass resolutions, as well as to specific sub-grid recipes.

\begin{acknowledgements}
This work has been carried out as part of the PHANGS collaboration. 
This work is based on observations made with the NASA/ESA/CSA JWST. The data were obtained from the Mikulski Archive for Space Telescopes at the Space Telescope Science Institute, which is operated by the Association of Universities for Research in Astronomy, Inc., under NASA contract NAS 5-03127 for JWST. These observations are associated with programs 2107 and 3707.

TGW gratefully acknowledges support from the UK ALMA Regional Centre (ARC) Node, which is supported by the Science and Technology Facilities Council grant number ST/Y004108/1.

ES and JDH acknowledge support from the European Research Council (ERC) under the Horizon Europe framework programme. This work is supported by an ERC Advanced Grant (Acronym: eCMZs, Number 101201230). Funded by the European Union. Views and opinions expressed are, however, those of the author(s) only and do not necessarily reflect those of the European Union or the European Research Council Executive Agency. Neither the European Union nor the granting authority can be held responsible for them. 

JGL acknowledges funding from the DLR (German Aerospace Agency) via grant 50 OR2401. 

MQ and MRG acknowledge support from the Spanish grant PID2022-138560NB-I00, funded by MCIN/AEI/10.13039/501100011033/FEDER, EU.

OE acknowledges funding from the Deutsche Forschungsgemeinschaft (DFG, German Research Foundation) -- project-ID 541068876.
Software: astropy \citep{Astropy+2013, Astropy+2018, Astropy+2022}. matplotlib \citep{Hunter2007}. numpy \citep{Harris+2020}.
Facilities: JWST.

\end{acknowledgements}

\bibliographystyle{aa}
\bibliography{library}

\begin{appendix}

\section{Portfolio of galaxies in the sample}
\label{sec:appthumb}
\begin{figure*}[h!]
\centering
\includegraphics[width=0.8\textwidth]{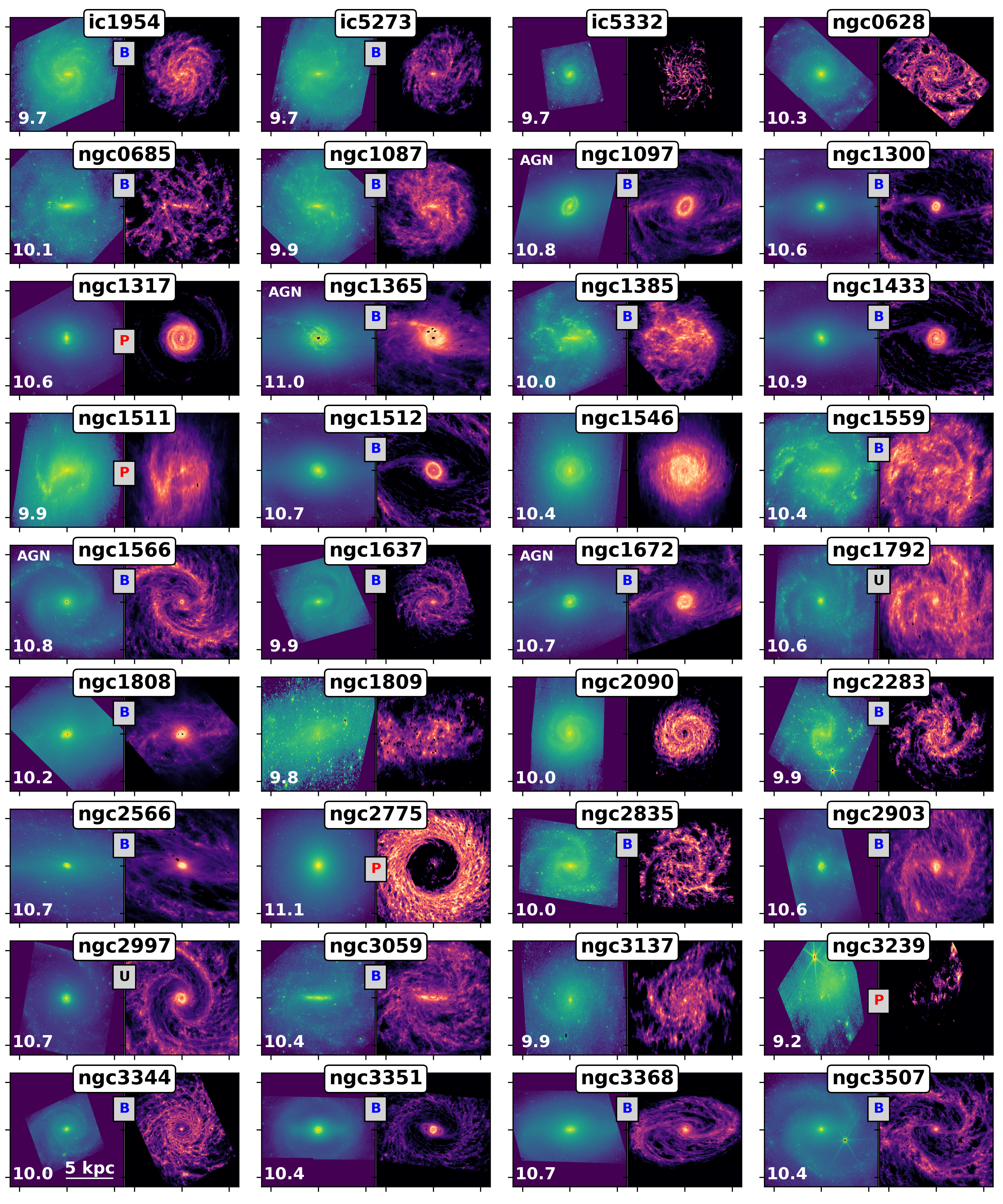}
\caption{Thumbnails showing the F300M and (continuum subtracted) F770W band JWST images within 6~kpc for all galaxies in the sample. Each pair of panels is associated with one target and includes the F300M (left) and F770W (right) images. Flags used throughout this paper are indicated with boxed letters. The top flag indicates if the galaxy has a bar (blue 'B'), is unbarred (no flag) or has an uncertain bar classification (black 'U'). Second row flags are P (red) for `perturbed', and E (red) for `Early-type'. Galaxies are ordered alphabetically, and their stellar masses (in log units) are indicated at the bottom left of each F300M panel.}
\label{fig:sample1}
\end{figure*}

\begin{figure*}[h!]
\centering
\includegraphics[width=0.8\textwidth]{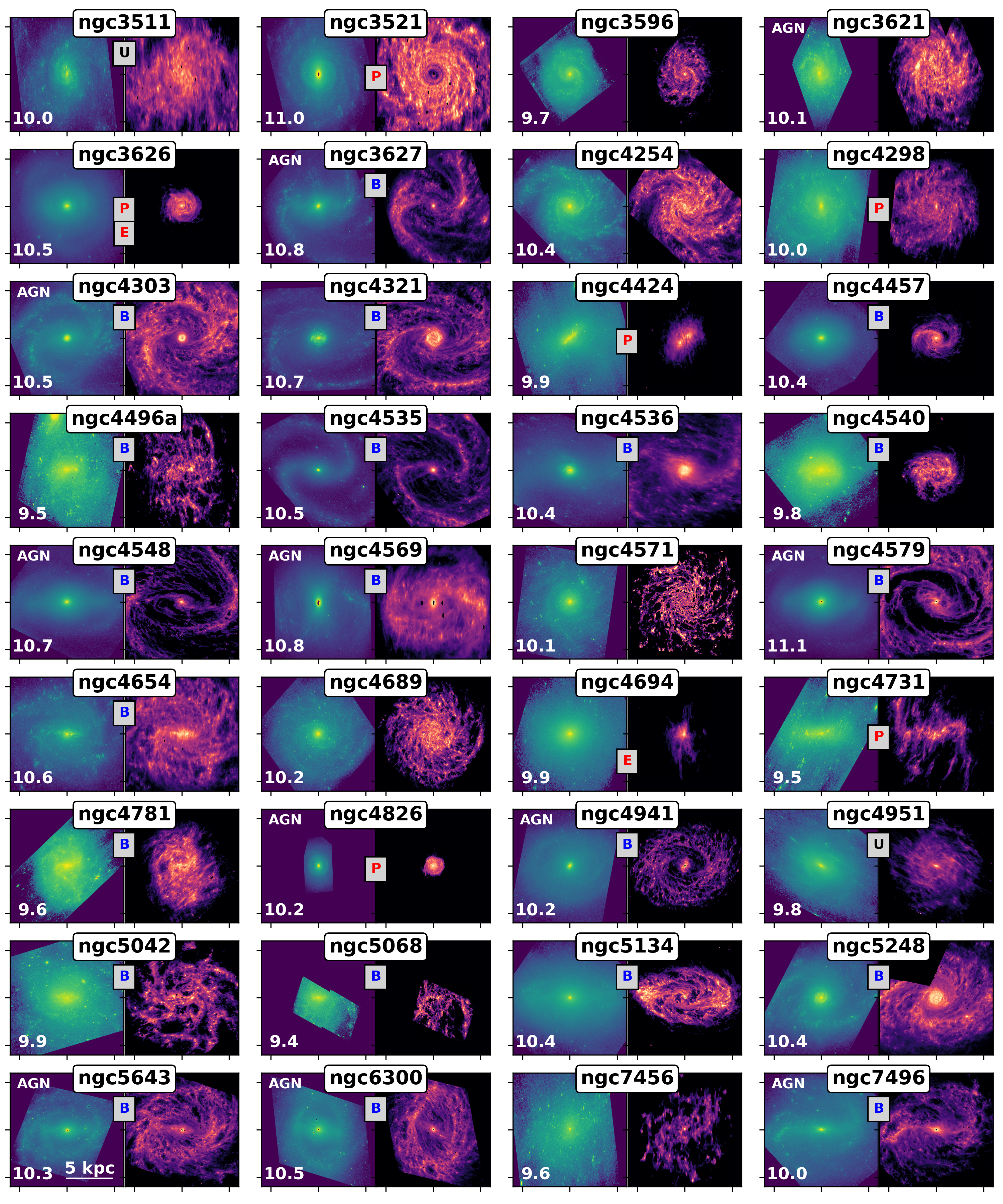}
\caption{Same as Fig.~\ref{fig:sample1}, for the second half of the sample (in alphabetical order).}
\label{fig:sample2}
\end{figure*}

In Figs.~\ref{fig:sample1} and \ref{fig:sample2}, we present the full set of JWST NIRCAM 7.7~$\mu$m- and MIRI 21~$\mu$m-band deprojected images for the 72 targets in our initial sample. The F770W band images have been continuum-subtracted using a weighted contribution from the F300M band.

\section{Fitting of radial surface density profiles}
\label{sec:appfit}
\begin{figure*}
\centering
\includegraphics[width=0.6\textwidth]{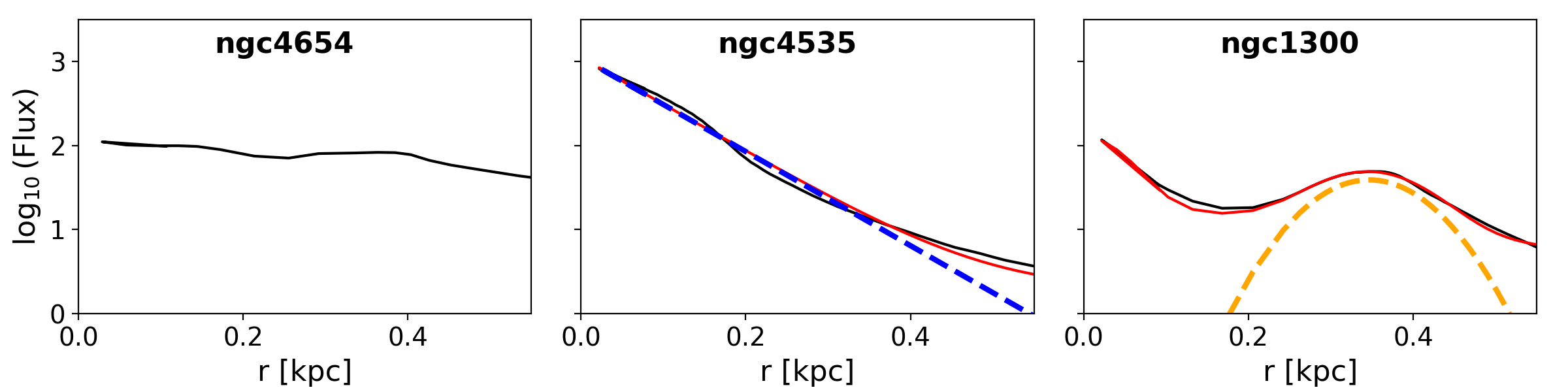}
\caption{Examples of fits for the F770W$_{\rm ss}$ deprojected radial surface brightness profiles. From left to right: NGC\,4564 with no detected central structure, NGC\,4535 with a central (exponential) disc-like structure, and NGC\,1300 with a ring at a radius of about 350~pc. The data is shown in black, while the global fit is in red and the individual components, central disc and ring, in blue and orange (dashed lines).}
\label{fig:radialfit}
\end{figure*}

\begin{figure*}
    \centering
    \includegraphics[width=0.95\textwidth]{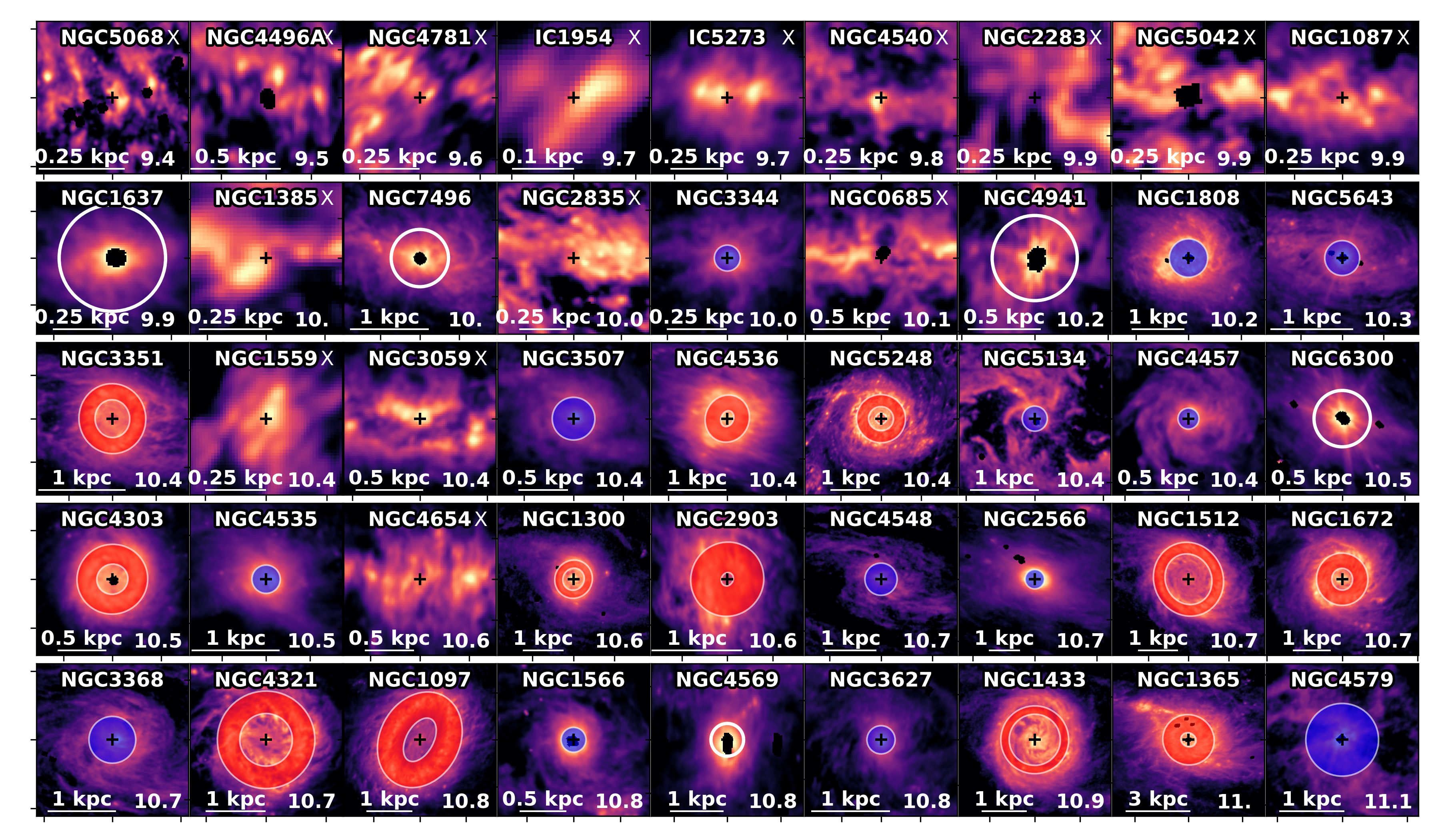}
    \caption{Zoom-in view of the F770W$_{\rm ss}$ JWST deprojected images for the 45 barred galaxies in our sample. The field of view in each panel has been adapted to cover a radius of 25\% of the bar size (an eight times zoom-in compared to Fig.~\ref{fig:thumbnails_770s}). White thick circles emphasise the presence of a bright central source that yields a saturated point spread function, which prevents a robust characterisation of the central structure. Blue-filled ellipses and orange annuli illustrate the presence of a central PAH emission reservoir and one (or two) rings, respectively. Panels with white crosses on the right of the galaxy name show targets without a detected central structure. Spatial scales are indicated via a bar in the lower left part of each panel.}
    \label{fig:rings770s}
\end{figure*}

\begin{figure*}
    \centering
    \includegraphics[width=0.95\textwidth]{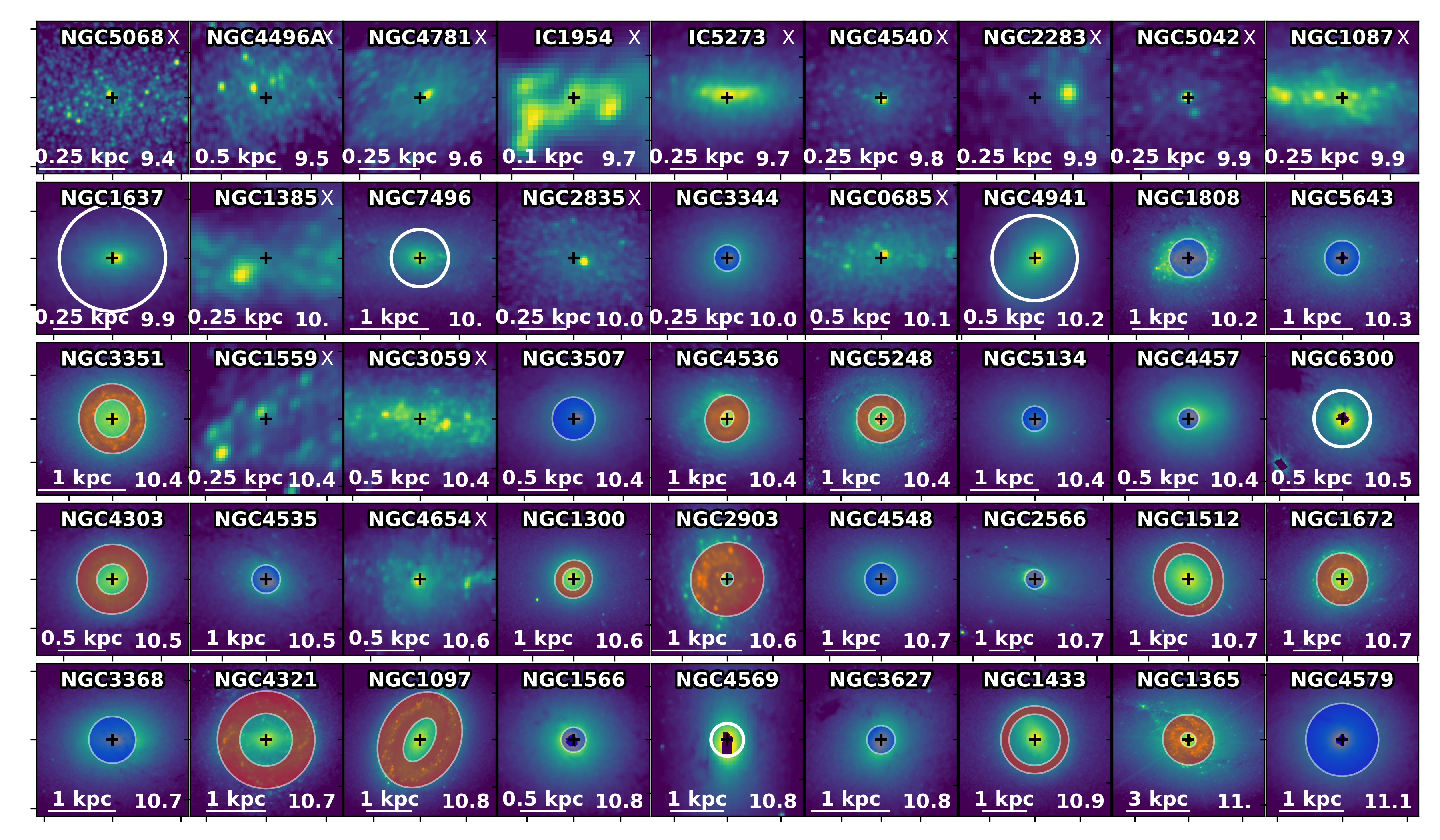}
    \caption{Same as Fig.~\ref{fig:rings770s} but for the F300M JWST band.}
    \label{fig:rings300}
\end{figure*}

In Fig.~\ref{fig:radialfit}, we present three examples of the outcome when fitting the radial deprojected surface density profiles in the JWST F770W$_{\rm ss}$ band. In the left panel, NGC\,4654 shows no detectable central feature, while NGC\,4535 exhibits a profile consistent with an inner exponential disc, and NGC\,1300 has a ring-like bump around 350~pc. We complement this information with the representation of the fitted structures onto the F770W$_{\rm ss}$ and F300M deprojected images in Figures~\ref{fig:rings770s} and \ref{fig:rings300}.


\section{Sample Table}
\label{sec:apptable}
\onecolumn
\begin{longtable}{lrrrrrrrrcccccc}
\caption{\label{tab:sample} Basic and deprojected characteristics of galaxies in the sample (see Sect.~\ref{sec:disc}).} \\
\hline\hline
Name & Dist & M$_{\star}$ & SFR & Inc & PA & PA$_{\rm b}$ & R$_{\rm b}^{\prime}$ & Type & CS & R & $\Delta R$ & $\epsilon$ & $\Theta$ \\
& Mpc & log$_{10}$\,\Msun\  & ${\rm M}_{\odot}$/yr & deg & deg & deg & kpc &  &  & pc & pc &  & deg \\
\hline
\endfirsthead
\caption{continued.}\\
\hline\hline
Name & Dist & M$_{\star}$ & SFR & Inc & PA & PA$_{\rm b}$ & R$_{\rm b}^{\prime}$ & Type & CS & R & $\Delta R$ & $\epsilon$ & $\Theta$ \\
& Mpc & log$_{10}$\,\Msun\  & ${\rm M}_{\odot}$/yr & deg & deg & deg & kpc &  &  & pc & pc &  & deg \\
\hline
\endhead
\hline
\endfoot
IC1954 & 12.8 & 9.7 & 0.4 & 57.1 & 63 & 90 & 0.5 & B & - & - & - & - & - \\
IC5273 & 14.2 & 9.7 & 0.5 & 52.0 & 234 & 39 & 1.4 & B & - & - & - & - & - \\
IC5332 & 9.0 & 9.7 & 0.4 & 26.9 & 74 & 0 &  & nB & - & - & - & - & - \\
NGC628 & 9.8 & 10.3 & 1.8 & 8.9 & 21 & 0 &  & nB & - & - & - & - & - \\
NGC685 & 19.9 & 10.1 & 0.4 & 23.0 & 101 & 55 & 2.0 & B & - & - & - & - & - \\
NGC1087 & 15.9 & 9.9 & 1.3 & 42.9 & 359 & 128 & 1.6 & B & - & - & - & - & - \\
NGC1097 & 13.6 & 10.8 & 4.7 & 48.6 & 122 & 141 & 6.6 & B & R & 647 & 272 & 0.20 & 152 \\
NGC1300 & 19.0 & 10.6 & 1.2 & 31.8 & 278 & 99 & 7.4 & B & R & 362 & 84 & 0.04 & 147 \\
NGC1317 & 19.1 & 10.6 & 0.5 & 23.2 & 222 & 150 & 3.9 & Pec & - & - & - & - & - \\
NGC1365 & 19.6 & 11.0 & 16.9 & 55.4 & 201 & 86 & 14.2 & B & R & 753 & 364 & 0.04 & 63 \\
NGC1385 & 17.2 & 10.0 & 2.1 & 44.0 & 181 & 80 & 1.0 & B & - & - & - & - & - \\
NGC1433 & 18.6 & 10.9 & 1.1 & 28.6 & 200 & 95 & 6.7 & B & R & 657 & 78 & 0.00 & 39 \\
NGC1511 & 15.3 & 9.9 & 2.3 & 72.7 & 297 & 0 &  & Pec & - & - & - & - & - \\
NGC1512 & 18.8 & 10.7 & 1.3 & 42.5 & 262 & 42 & 7.6 & B & R & 744 & 127 & 0.07 & 24 \\
NGC1546 & 17.7 & 10.4 & 0.8 & 70.3 & 148 & 0 &  & nB & - & - & - & - & - \\
NGC1559 & 19.4 & 10.4 & 3.8 & 55.0 & 244 & 82 & 1.3 & B & - & - & - & - & - \\
NGC1566 & 17.7 & 10.8 & 4.5 & 29.5 & 215 & 177 & 3.2 & B & D & 66 & - & - & - \\
NGC1637 & 11.7 & 9.9 & 0.6 & 31.1 & 21 & 85 & 1.3 & B & S & 88 & - & - & - \\
NGC1672 & 19.4 & 10.7 & 7.6 & 42.6 & 134 & 96 & 8.1 & B & R & 482 & 173 & 0.02 & 171 \\
NGC1792 & 16.2 & 10.6 & 3.7 & 65.1 & 319 & 0 &  & Unc & - & - & - & - & - \\
NGC1808 & 12.8 & 10.2 & 0.0 & 49.0 & -53 & 160 & 5.8 & B & D & 182 & - & - & - \\
NGC1809 & 20.0 & 9.8 & 5.7 & 57.6 & 138 & 0 &  & nB & - & - & - & - & - \\
NGC2090 & 11.8 & 10.0 & 0.4 & 64.5 & 192 & 0 &  & nB & - & - & - & - & - \\
NGC2283 & 13.7 & 9.9 & 0.5 & 43.7 & -4 & 173 & 0.8 & B & - & - & - & - & - \\
NGC2566 & 23.4 & 10.7 & 8.7 & 48.5 & 312 & 66 & 9.8 & B & D & 155 & - & - & - \\
NGC2775 & 23.1 & 11.1 & 0.9 & 41.2 & 156 & 0 &  & Pec & - & - & - & - & - \\
NGC2835 & 12.2 & 10.0 & 1.2 & 41.3 & 1 & 115 & 1.6 & B & - & - & - & - & - \\
NGC2903 & 10.0 & 10.6 & 3.1 & 66.8 & 204 & 28 & 3.4 & B & R & 236 & 145 & 0.03 & 28 \\
NGC2997 & 14.1 & 10.7 & 4.4 & 33.0 & 108 & 100 & 0.8 & Unc & - & - & - & - & - \\
NGC3059 & 20.2 & 10.4 & 2.4 & 29.4 & -15 & 40 & 2.3 & B & - & - & - & - & - \\
NGC3137 & 16.4 & 9.9 & 0.5 & 70.3 & -0 & 0 &  & nB & - & - & - & - & - \\
NGC3239 & 10.9 & 9.2 & 0.4 & 60.3 & 73 & 0 &  & Pec & - & - & - & - & - \\
NGC3344 & 9.8 & 10.0 & 0.8 & 27.0 & 143 & 178 & 1.3 & B & D & 27 & - & - & - \\
NGC3351 & 10.0 & 10.4 & 1.3 & 45.1 & 193 & 112 & 3.5 & B & R & 298 & 80 & 0.05 & 173 \\
NGC3368 & 11.2 & 10.7 & 0.7 & 51.1 & 5 & 124 & 4.4 & B & D & 171 & - & - & - \\
NGC3507 & 23.5 & 10.4 & 1.0 & 21.7 & 56 & 112 & 3.1 & B & D & 107 & - & - & - \\
NGC3511 & 13.9 & 10.0 & 0.8 & 75.1 & 257 & 70 & 1.3 & Unc & - & - & - & - & - \\
NGC3521 & 13.2 & 11.0 & 3.7 & 68.8 & 343 & 0 &  & Pec & - & - & - & - & - \\
NGC3596 & 11.3 & 9.7 & 0.3 & 25.1 & 78 & 0 &  & nB & - & - & - & - & - \\
NGC3621 & 7.1 & 10.1 & 1.0 & 65.8 & 344 & 0 &  & nB & - & - & - & - & - \\
NGC3626 & 20.0 & 10.5 & 0.2 & 46.6 & 165 & 167 & 1.9 & Pec & - & - & - & - & - \\
NGC3627 & 11.3 & 10.8 & 3.8 & 57.3 & 173 & 160 & 3.9 & B & D & 89 & - & - & - \\
NGC4254 & 13.1 & 10.4 & 3.1 & 34.4 & 68 & 0 &  & nB & - & - & - & - & - \\
NGC4298 & 14.9 & 10.0 & 0.5 & 59.2 & 314 & 0 &  & Pec & - & - & - & - & - \\
NGC4303 & 17.0 & 10.5 & 5.3 & 23.5 & 312 & 178 & 3.1 & B & R & 257 & 87 & 0.02 & 112 \\
NGC4321 & 15.2 & 10.7 & 3.6 & 38.5 & 156 & 108 & 5.1 & B & R & 625 & 160 & 0.00 & 77 \\
NGC4424 & 16.2 & 9.9 & 0.3 & 58.2 & 88 & 0 &  & Pec & - & - & - & - & - \\
NGC4457 & 15.1 & 10.4 & 0.3 & 17.4 & 79 & 65 & 2.4 & B & D & 41 & - & - & - \\
NGC4496A & 14.9 & 9.5 & 0.6 & 53.8 & 51 & 49 & 1.7 & B & - & - & - & - & - \\
NGC4535 & 15.8 & 10.5 & 2.2 & 44.7 & 180 & 42 & 3.5 & B & D & 80 & - & - & - \\
NGC4536 & 16.2 & 10.4 & 3.4 & 66.0 & 306 & 77 & 5.2 & B & R & 249 & 118 & 0.08 & 158 \\
NGC4540 & 15.8 & 9.8 & 0.2 & 28.7 & 13 & 62 & 1.5 & B & - & - & - & - & - \\
NGC4548 & 16.2 & 10.7 & 0.5 & 38.3 & 138 & 61 & 5.9 & B & D & 156 & - & - & - \\
NGC4569 & 15.8 & 10.8 & 1.3 & 70.0 & 18 & 18 & 5.7 & B & S & 86 & - & - & - \\
NGC4571 & 14.9 & 10.1 & 0.3 & 32.7 & 218 & 0 &  & nB & - & - & - & - & - \\
NGC4579 & 21.0 & 11.1 & 2.2 & 40.2 & 91 & 53 & 4.7 & B & D & 279 & - & - & - \\
NGC4654 & 22.0 & 10.6 & 3.8 & 55.6 & 123 & 121 & 3.1 & B & - & - & - & - & - \\
NGC4689 & 15.0 & 10.2 & 0.4 & 38.7 & 164 & 0 &  & nB & - & - & - & - & - \\
NGC4694 & 15.8 & 9.9 & 0.2 & 60.7 & 143 & 0 &  & Etg & - & - & - & - & - \\
NGC4731 & 13.3 & 9.5 & 0.6 & 64.0 & 114 & 127 & 3.7 & Pec & - & - & - & - & - \\
NGC4781 & 11.3 & 9.6 & 0.5 & 59.0 & 290 & 82 & 1.3 & B & - & - & - & - & - \\
NGC4826 & 4.4 & 10.2 & 0.2 & 59.1 & 294 & 0 &  & Pec & - & - & - & - & - \\
NGC4941 & 15.0 & 10.2 & 0.4 & 53.4 & 202 & 16 & 2.1 & B & S & 148 & - & - & - \\
NGC4951 & 15.0 & 9.8 & 0.4 & 70.2 & 91 & 0 &  & Unc & - & - & - & - & - \\
NGC5042 & 16.8 & 9.9 & 0.6 & 49.4 & 191 & 75 & 1.6 & B & - & - & - & - & - \\
NGC5068 & 5.2 & 9.4 & 0.3 & 35.7 & 342 & 153 & 0.9 & B & - & - & - & - & - \\
NGC5134 & 19.9 & 10.4 & 0.5 & 22.7 & 312 & 153 & 4.4 & B & D & 91 & - & - & - \\
NGC5248 & 14.9 & 10.4 & 2.3 & 47.4 & 109 & 135 & 7.6 & B & R & 91 / 453 & 33 / 128 & 0.00 / 0.02 & 47 / 80 \\
NGC5643 & 12.7 & 10.3 & 2.6 & 29.9 & 319 & 84 & 3.7 & B & D & 105 & - & - & - \\
NGC6300 & 11.6 & 10.5 & 1.9 & 49.6 & 105 & 71 & 2.4 & B & S & 85 & - & - & - \\
NGC7456 & 15.7 & 9.6 & 0.4 & 67.3 & 16 & 0 &  & nB & - & - & - & - & - \\
NGC7496 & 18.7 & 10.0 & 2.3 & 35.9 & 194 & 147 & 3.9 & B & S & 127 & - & - & - \\
\hline
\end{longtable}
\tablefoot{Columns are: (1) target name; (2) distance; (3) stellar mass; (4) star formation rate; (5) Inclination (6) position angle of the line of nodes; (7) position angle of the bar; (8) Deprojected radius of the bar; (9) host type such as: barred (B), not barred (nB), peculiar (Pec), early-type (Etg), unclassified (Unc); (10) central structure (CS) such as a central disc (D), ring (R) or saturated centre (S); (11) radius of the structure; (12) characteristic width of the structure: (13) ellipticity of the structure and (14) angle of the central structure with respect to the deprojected axis of the bar. Most physical properties in this table have been compiled from \cite{Leroy+2021}, and the morphological and intrinsic properties (inclination, kinematic position angle) from \cite{Querejeta+2021, Lang+2020}. See also Sect.~\ref{sec:subsample} for specific changes to the bar classification, and Sect.~\ref{sec:reservoir} for the characteristics of the central structures.}
\end{appendix}

\end{document}